\documentclass[a4paper,10pt,aps,prd,floatfix,notitlepage,showpacs]{revtex4-1}
\usepackage{amsmath,amssymb,amsfonts}

\usepackage{amsmath}
\usepackage{graphicx,color,float}
\usepackage{subfig}

\usepackage{epstopdf,graphics} 

\newcommand{\ba}{\begin{eqnarray}}
	\newcommand{\ea}{\end{eqnarray}}

\def\ben{\begin{equation}}
	\def\een{\end{equation}}
\def\bea{\begin{eqnarray}}
	\def\eea{\end{eqnarray}}
\def\be{\begin{equation}}
	\def\ee{\end{equation}}

\def\nn{\nonumber}

\input amssym.def
\input amssym.tex

\def\fft#1#2{{\frac{#1}{#2}}}

\usepackage[T1]{fontenc} 

\begin{document}

\title{\boldmath Thermodynamic and Holographic Information Dual to Volume }



\author{Davood Momeni}
\email{davoodmomeni78@gmail.com}
\affiliation{Department of Physics, College of Science, Sultan Qaboos University,
	\\P.O. Box 36, P.C. 123, Al-Khodh, \\Muscat, Sultanate of Oman}

\author{Mir Faizal}
\email{mirfaizalmir@googlemail.com}
\affiliation{Irving K. Barber School of Arts and Sciences, University of British Columbia – Okanagan, Kelowna, British Columbia V1V 1V7, Canada}
\affiliation{Department of Physics and Astronomy, University of Lethbridge, Lethbridge, Alberta, T1K 3M4, Canada.}

\author{Salwa Alsaleh}
\email{salwams@ksu.edu.sa}
\affiliation{Department of Physics and Astronomy, College of Science
King Saud University. Riyadh 11451 , Saudi Arabia}

\author{Lina Alasfar}
\email{ linana343@outlook.com}
\affiliation{Laboratoire de Physique Corpusculaire de Clermont-Ferrand, Universit\'{e} Blaise Pascal, 24 Avenue des Landais F-63177 Aubi\`{e}re Cedex, Francea}

\author{Aizhan Myrzakul}
\email{a.r.myrzakul@gmail.com}
\affiliation{Eurasian International Center for Theoretical Physics and
	Department of General \& Theoretical Physics, Eurasian National
	University, Astana 010008, Kazakhstan}

\author{Ratbay Myrzakulov}
\email{rmyrzakulov@gmail.com}
\affiliation{Eurasian International Center for Theoretical Physics and
	Department of General \& Theoretical Physics, Eurasian National
	University, Astana 010008, Kazakhstan}

\begin{abstract}
In this paper, we will analyze the connection between 
the fidelity susceptibility, the holographic 
complexity and the thermodynamic volume. We will regularize the 
fidelity susceptibility and the holographic 
complexity by subtracting 
the  contribution of  the background AdS spacetime from the deformation 
of the AdS spacetime. 
It will be demonstrated that this regularized 
fidelity susceptibility  has the same behavior as the thermodynamic volume 
and that the regularized complexity 
has a very different behavior. As the information dual to different 
volumes in the bulk would be measured by the fidelity susceptibility and the holographic 
complexity, this paper will establish a connection between thermodynamics and information 
dual to a volume.
\end{abstract}
\maketitle


\section{Introduction}
It has been observed 
 from various studies done in different branches of physics that the laws of physics 
 are related to  the ability of an 
observer to process 
  relevant information \cite{info, info2}. Thus, it seems to indicate 
  that the laws of physics are information theoretical processes. 
As the  informational theoretical process deal with the processing of information, 
it is important to measure the loss of information during such a process. 
It is possible to measure this loss of information using the concept of entropy, 
and so that the entropy is a very important quantity in information theory. 
However, as laws of physics are also information theoretical processes, 
entropy is an important physical quantity.
In fact, using 
the Jacobson formalism, it is possible to obtain the geometric structure of spacetime  by 
assuming a certain scaling behavior of the maximum entropy of a region of 
space 
 \cite{z12j, jz12}.
 This scaling behavior of maximum entropy has been motivated from the   
 the holographic principle
 \cite{1, 2}. 
 The holographic principle states that the number of degrees of freedom 
 in any region of space is equal to the number of degrees of freedom 
 on the boundary of that region of space. The holographic principle is also the 
 basis of the  the   AdS/CFT conjecture 
 \cite{M:1997}. 
 The AdS/CFT correspondence states that the 
supergravity/string theory 
in the bulk of an 
AdS spacetime is dual to the superconformal field theory on its boundary. 
The AdS/CFT correspondence  has been used to 
quantify the  concept of 
entanglement  in conformal field theory. This is because 
 the AdS/CFT correspondence can be used 
to holographically  calculate  
quantum entanglement entropy of a conformal field theory from the bulk AdS spacetime  \cite{4,5,Momeni:2014qta}.
This is done by defining   $\gamma_{A}$ as
   the  $(d-1)$-minimal surface extended for a 
   subsystem $A$  with  boundary $\partial A$. 
   The  holographic entanglement 
entropy for such a subsystem can be  expressed in terms of the
 the gravitational constant $G_{d+1}$ and 
area of the 
minimal surface $Area[{A}]$ as
  \cite{6, 6a}
\begin{equation} \label{HEE}
\mbox{Entropy}_{A}=\frac{Area [{A}](\gamma _{A})}{4G_{d+1}}\,.
\end{equation}
Since the entropy is related to the loss of information, it can 
be calculated holographically from the area of a minimal surface. 
However, it is not only important to understand the loss of information 
during an information theoretical 
process, but it is also important to understand the difficulty 
to process that information during such a process. 
This can be quantified using the concept of complexity. 
As laws of physics are informational theoretical processes, it 
is expected that complexity will become an important physical 
quantity, and the fundamental laws of physics will be expressed 
in terms of complexity.  It may be noted that complexity 
has already been used to understand the behavior of certain condensed matter
 systems \cite{c1, c2,Momeni:2017kbs}. It has also been used for analyzing 
 molecular physics \cite{comp1,Momeni:2017ibg}. In fact, even 
 quantum computational systems have been studied using the concept 
 of complexity \cite{comp2}.  The studies done on black hole information indicate that 
 the information might not be lost, but it would be left in such a state that 
 it would not be effectively possible to recover it from that state \cite{hawk}. 
 This indicates that complexity might be an important quantity that can be used 
 to understand the black hole information paradox. 
It has been proposed that the complexity can also be holographically 
calculated from the bulk AdS spacetime \cite{Susskind:2014rva1,Susskind:2014rva2}.
In fact, it has been proposed that the 
 complexity would be dual to a volume $V$ in the bulk AdS spacetime and hence it can be defined as follows \cite{Stanford:2014jda,Momeni:2016ekm}
\begin{equation}\label{HCC}
\mbox{Complexity} = \frac{V }{8\pi R G_{d+1}},
\end{equation}
where $R$ is  the  radius of the curvature.  
There are various different ways to define a radius in AdS, and so we have different proposals for 
complexity.  It is possible to use the same minimal surface which was used to calculate 
the holographic entanglement entropy, and define this volume as the volume enclosed 
by such a surface   $V = V(\gamma)$ \cite{Alishahiha:2015rta}. However, this quantify diverges, 
and so we will regularize it by subtracting the contribution of a 
background  AdS  $V(\gamma)_{AdS}$ from 
the deformation of  AdS $V(\gamma)_{AdS}$, and define 
\begin{equation}\label{HCCC}
 \Delta \mathcal{C} = \frac{V(\gamma)_{DAdS} -  V(\gamma)_{AdS}}{8\pi R G_{d+1}}.
\end{equation}
This quantity will be finite, and we shall call this quantity as the 
holographic complexity. It is also possible to define 
the volume in the bulk as the  
  maximal volume in the AdS which
ends on the time slice at the AdS boundary, $V = V(\Sigma_{max})$ \cite{MIyaji:2015mia}. 
This would again leads to divergences, and so we will again need to regularize it. 
The later can be achieved by subtracting  the contribution of a 
background  AdS  $V(\gamma)_{AdS}$ from 
the deformation of  the AdS $V(\gamma)_{AdS}$, and defining 
\begin{equation}\label{HC}
\Delta \chi_F = \frac{V(\Sigma_{max})_{DAdS} -  V(\Sigma_{max})_{AdS}}{8\pi R G_{d+1}}.
\end{equation}
This quantity will be finite, and it will correspond to the 
fidelity susceptibility  of a boundary field theory \cite{r6,r7, r8}. 
So, we shall call this quantity as the fidelity susceptibility  even in the bulk. 
It has recently been proposed that the fidelity susceptibility has the same behavior 
as the thermodynamic volume in the extended phase space \cite{paper}. 
The   cosmological constant is treated as the thermodynamics
pressure in extended phase space, and it is possible to define a thermodynamic volume 
conjugate to this pressure \cite{ext2}-\cite{Zhang:2017nth}.
In this paper, we will analyze the relation between the thermodynamic 
volume, fidelity susceptibility and holographic complexity for different black hole 
solutions. It will be observed that for all these different black hole solutions, 
the  thermodynamic 
volume has the same behavior as the  fidelity susceptibility. Thus, the 
recently observed behavior for  a specific black hole solution \cite{paper}, 
seems to be a universal behavior of  thermodynamic 
volume and  fidelity susceptibility. Furthermore, it will be observed that the 
holographic complexity is different for all these cases.  The information dual to a volume in the bulk is measured by the regularized fidelity
 susceptibility and the regularized holographic 
 complexity \cite{faiz,Ben-Ami:2016qex,Roy:2017kha,Gan:2017qkz}, so this paper establishes a connection between the information dual to a volume  and thermodynamics. We would like to point out  that the original fidelity susceptibility and the original 
holographic complexity 
were divergences, and we have regularized them. Therefore, when we refer to the
fidelity susceptibility and the  
holographic complexity, we are actually referring to these regularized  fidelity susceptibility and holographic complexity. 

\section{Schwarzschild Anti-de Sitter black holes}\label{Sads}
In this section, we consider black holes in AdS space as deformations that correspond to excited states on the CFT boundary. Then we calculate the temprature, entropy, and heat capacity. After that, we calculate $ \Delta \mathcal C$ and $ \Delta \chi_F$  by first calculating the  volume of minimal surface of the AdS slice, and the maximal volume of that slice  \cite{paper}.
 We begin our study with a simple case where the geometry is a topological Schwarzschild Anti de Sitter (SAdS) black hole, whose metric is represented by the following line-element, 
\be
ds^2=-f(r)dt^2+\fft{dr^2}{f(r)}\,+r^2d\Omega^2\,,\label{metric1}
\ee
where $d\Omega^2=d\theta^2+\frac{1}{k}\sin^2(\sqrt{k}\theta)d\phi^2$ with $k=\{-1, 0, 1\}$. Additionally, the metric function $f(r)$ is given by
\be
f(r)=1-\frac{2M}{r}+\frac{r^2}{l^2}\,.
\ee
By assuming that $r_{+}$ is the black hole event horizon, i.e.  $f(r_{+})=0$, we can introduce a small parameter  $\epsilon=\frac{Ml^2}{r_{+}^3}\ll 1$ to rewrite the above function yielding 
\be
f=1+\frac {1}{l^2} \left(r^2 - \frac {2\epsilon\,r_{+}^3}{r} \right)\,.
\ee  
The mass and the volume of the sphere enclosed by the black hole are 
\be
M=\frac{r_+}{2}\left(1+\frac{r_{+}^2}{l^2}\right)\,,\label{M} \qquad
V=\frac{4}{3}\pi r_{+}^3\,. \qquad
\ee
The temperature and specific heat of the black hole are respectively
\begin{eqnarray}
T &=&\frac{1}{4\pi r_{+}l^2}\left(l^2+3r_{+}^2 \right)\,,\\
c_p&=&\left(\frac{\partial M}{\partial T}\right)_p=\frac{\frac{\partial M}{\partial r_{+}}}{\frac{\partial T}{\partial r_{+}}}=\frac{\frac{\partial}{\partial r_{+}}\left(\frac{r_{+}}{2}+\frac{r_{+}^3}{2l^2}\right)}{\frac{\partial}{\partial r_{+}}\left(\frac{1}{4\pi l^2 r_+}\left(l^2+3r_{+}^2\right)\right)}=2\pi r_{+}^2\left(\frac{3r_{+}^2+l^2}{3r_{+}^2-l^2}\right)\,.
\end{eqnarray}
Now, let us compute the area of the minimal surface $\gamma$ which is parametrized by $r=r(\theta)$ given by a time slice $t=0$ in the line-element, 
\be
ds^2\mid_{t=0}=r^2\sin^2{\theta}d\phi^2+\left (r^2+\frac{(\frac{dr}{d\theta})^2}{f(r)}\right)d\theta^2\,.
\ee 
Hence, the minimal area can be written as follows
\be \label{area1}
\mbox{Area}
=\int_0^{2\pi}d\phi \int_{0}^{\theta_0} L(\theta)d\theta\,,
\ee
where $L(\theta)=r \sin \theta \sqrt {r^2+\frac{(\frac{dr}{d\theta})^2}{f(r)}}$ and $\theta_0$ is the upper boundary on the entangled domain. Therefore, the total area will be given by
\be
A=2(2\pi)\int_0^{\theta_0} L(\theta)d\theta\,.\label{areatotal}
\ee
Now, we will assume the following boundary conditions imposed on  minimal surface
\be
	r'(0)=0\,,	r(0)=\rho\,,
\ee
where $\rho$  is the turning point of the solution $r(\theta)$ and prime denotes differentiation with respect to $\theta$. The corresponding solution of the Euler-Lagrange equation for this Lagrangian density $L$ and the above boundary condition is given by
\begin{eqnarray}
r( \theta) &=&\rho-\frac{1}{2}{\frac { \left( -{\rho}^{3}-\rho\,{
l}^{2}+2\,\epsilon\,r_+^3 \right) {\theta}^{2}}{{l}^{2}}} \nonumber\\
&&+\frac{1}{48}{\frac { \left( 18\,{\rho}^{6}+28\,{\rho}^{4}{l}^{2}-45\,\epsilon\,r_+^
3{\rho}^{3}+10\,{\rho}^{2}{l}^{4}-29\,\rho\,{l}^{2}\epsilon\,r_+^3
+18\,{\epsilon}^{2}r_+^6 \right) {\theta}^{4}}{\rho\,{l}^{4}}}\,.
\end{eqnarray}
Additionally, we obtain that $L=-3( {\rho}r_+^3{\theta}^{3})/{l}^{2}$ and hence, from (\ref{areatotal}), the total area can be expressed as 
\be
A=-3\pi{\frac {{csgn} \left( \rho \right) \rho\,r_+^3{\theta_0^4}}{{l}^{2}}}\,.
\ee
where 
\be
csgn(\rho)=
\begin{cases}
1\,,& \rho>0
\\
-1\,,& \rho<0
\end{cases}\,.
\ee
Consequently, the difference of entanglement entropy between pure AdS and SAdS  with $\rho>0$ is
\be
\Delta S=-\frac{3}{16}\frac{\rho^3 r_{+}^3 \theta_0^{4}}{Gl^2},
\ee
where $G$ is a gravitational constant.
\par
\begin{figure}

\includegraphics[scale=0.5]{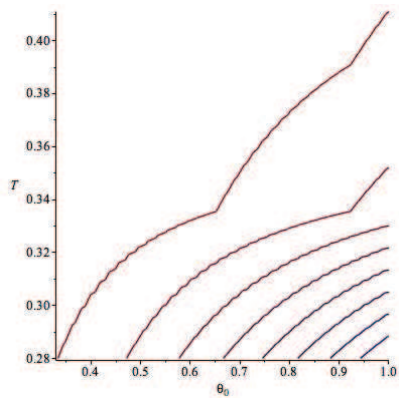}
\centering
\label{Fig1}
\caption{Figures show contour plots of $T=\mbox{constant}$ in $V$  as functions of $\theta_0$. Different lines in each graph show a specific temperature. }
\end{figure}

In order to compute the holographic complexity, we need to evaluate  the volume of the bulk  enclosed by the same  surface used in entanglement entropy. This volume is defined as follows
\be
V=2\pi\int_0^{\theta_0}d\theta\int_{r_{+}}^{r(\theta)}\frac{r^2dr}{\sqrt{f(r)}}\,.\label{V}
\ee
In order to solve the integral in the above expression, we can expand the integrand in Taylor series of the parameter $\epsilon$ as
\be
\frac{r^2}{\sqrt{f(r)}}
={\frac {r^2}{\sqrt {{\frac {{l}^{2}+{r}^{2}}{{l}^{2}}}}}}+{\frac{r r_+
^3}{\sqrt {{\frac {{l}^{2}+{r}^{2}}{{l}^{2}}}}\left( {l}
^{2}+{r}^{2} \right)}} \epsilon+\mathcal{O}( {\epsilon}^{2} )\,.
\ee
Then, by leaving only linear terms in $\epsilon$ and expanding in $\theta$ up to second order we find that the volume is 
\be
\Delta V
=\frac{1}{2}\left( -{\frac {{r_+^3}l}{\sqrt {{l}^{2}+\rho^2}}}+{\frac {{
r_+^3}l}{\sqrt {{l}^{2}+{r_+^2}}}} \right) {\theta_0^2}\,.
\ee
Now, using the above volume, we can finally compute the holographic complexity, which is
\be
\Delta \mathcal{C} =\frac{\Delta V}{8\pi l}
=\frac{1}{2}\left( -{\frac {{r_+^3}l}{\sqrt {{l}^{2}+\rho^2}}}+{\frac {{
r_+^3}l}{\sqrt {{l}^{2}+{r_+^2}}}} \right) {\theta_0^2}\,,\label{VV}
\ee
where $\Delta V=V_{SAdS}-V_{AdS}$. 
Additionally, the complexity pressure $P_c$ can be defined using the ordinary thermodynamic relation $P=-\frac{\partial M}{\partial V_c}$, yielding
\be
P_c=-\frac{\frac{\partial M}{\partial r_{+}}}{\frac{\partial V_c}{\partial r_{+}}}=-{\frac { \left( 1+3r_+^{2} \right) \sqrt {1+\rho^2} \left( 1+r_+^2 \right) ^{3/2}}{{r_+^2} \left( -3\sqrt {1+r_+^{2}}r_+^{2}-3
\sqrt {1+r_+^{2}}+2r_+^{2}\sqrt {1+{\rho}^{2}}+3\sqrt {1+{\rho}^{2}} \right) {\theta_0^2}}}\,,\label{P}
\ee
where the mass of black hole is given in (\ref{M}). Moreover, from (\ref{M}), we can express the event horizon of the black hole in terms of the temperature, obtaining
\be
r_{+}=\frac{2}{3}\pi T\pm\frac{1}{3}\sqrt {4\pi^2T^2-3}\,.\label{rplus}
\ee
It is important to remark that we will choose the root with the minus sign since only this root has a physical meaning for the case where $k>1$.
Now, by using the above expression, we can rewrite the volume and the complexity pressure given by (\ref{VV}) and (\ref{P}) in terms of the temperature.  Note that the explicit expressions are found in the appendix \ref{appendixA}.
 Since both functions 
$P,V$ are functions of the temperature $T$ and the parameter $\theta_0$, it is illustrative to plot these functions 
 as contour plots. These figures are displayed in Fig.~\ref{Fig1,fig1a}. In each graph we fixed the temperature $T$ and consequently each function $P,V$ are defined as a single variable function of $\theta_0$. Since $V$ and $P$ depend on $\theta_0$, it is also possible to rewrite $P=P(T,V)$ as an equation of state if we eliminate $\theta_0$ among these two expressions for $P$ and $V$. 
and rewrite the complexity volume and pressure in terms of temperature. 
\begin{figure}
	\centering
\includegraphics[scale=0.5]{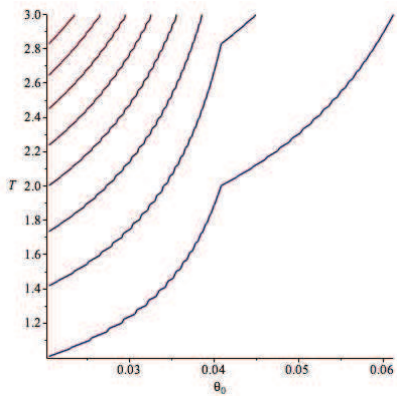}
\label{Fig1a}
\caption{Figures show contour plots of $T=\mbox{constant}$ in  $P$ as functions of $\theta_0$. Different lines in each graph show a specific temperature. }
\end{figure}
\\
Another volume dual to the thermodynamical volume in the bulk could be the maximum volume which is proportional to the 
fidelity susceptibility in the dual CFT part. The later is defined as follows
\be
\Delta \chi_F =\frac{V_{max}}{8\pi l G}.
\ee 
where $l$ is the AdS radius and $G$ the Newtonian constant. To evaluate $V_{max}$, we should subtract the pure AdS background portion. We can expand in series the metric function because $\epsilon=0$ corresponds to the AdS background. 
\be
\frac{1}{\sqrt{f}}=\frac{1}{\sqrt{f_0+\epsilon \delta f}}\simeq \frac{1}{\sqrt{f_0}}\left(1-\frac{\epsilon \delta f}{2 f_0}\right).
\ee
After a simple algebraic manipulation we find
\be
V_{Fid}=V_{AdS}+\epsilon\Delta V_{Fid} = \int_{r_+}^{r_\infty} \frac{r dr}{\sqrt{f_0}}\left(1-\frac{\epsilon\delta f}{2f_0}\right)\simeq \infty-\frac{\epsilon}{2}\int_{r_+}^{r_\infty} \frac{r\delta f}{{f_0}^{3/2}}dr\,.
\ee
Using the fact that $f_0=1+\frac{r^2}{l^2}$ (the metric of pure AdS) and $\delta f=-\frac{2r_{+}^3}{r}$, the corresponding volume is
\begin{eqnarray}
\Delta V_{Fid}&=&-\frac{r_+^3}{8(1
	+2{r_+^2}+{r_+^4})}\Big [  10r_{+}+6\arctan \left( r_+ \right) +6
\arctan \left( r_+ \right) {r_+^4}\nonumber\\
&&+6{r_+^3}+12\arctan \left( r_+
 \right) {r_+^2}-3\pi -6\pi {r_+^2}-3\pi {r_+^4}\Big],
\end{eqnarray}
and the pressure reads
\begin{eqnarray}
P_{Fid}&=&-\frac{\partial M}{\partial V_{Fid}}=-\frac{\frac{\partial M}{\partial r_{+}}}{\frac{\partial V_{Fid}}{\partial r_{+}}}\nonumber\\
&=&-\frac{\frac{4}{3}\left( \frac{1}{3}+{r_+^2} \right)  \left( 1+{r_+^2} \right) ^{3} }{r_+^2\left( -2\left( 1+{r_+^2} \right) ^{3}\arctan \left( r_+ \right) +
\pi {r_+^6}-\frac{16}{3}{r_+^3}-2{r_+^5}+3\pi {r_+^2}+3\pi {r_+^
4}+\pi -{\frac {46}{9}}r_+ \right)}.\nonumber\\
\end{eqnarray}
Note that $r_{+}$  is given by Eq.~(\ref{rplus}). 
The parametric plot of $p=p(V, T)$  is depicted in Fig.~\ref{fig1b}.
\begin{figure}
\centering
\includegraphics[scale=0.5]{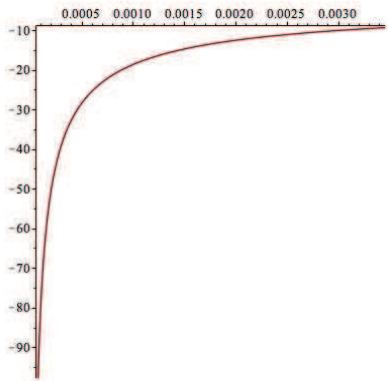}
\caption{This is a parametric plot for fidelity pressure and volume. This graph shows a fidelity based equation of state as an attempt to find holographic version of thermodynamics.}
\label{fig1b}
\end{figure}
Here the horizontal line corresponds to $V$ and the vertical line to $p$.
\begin{figure}	
	\centering
		\includegraphics[scale=0.35]{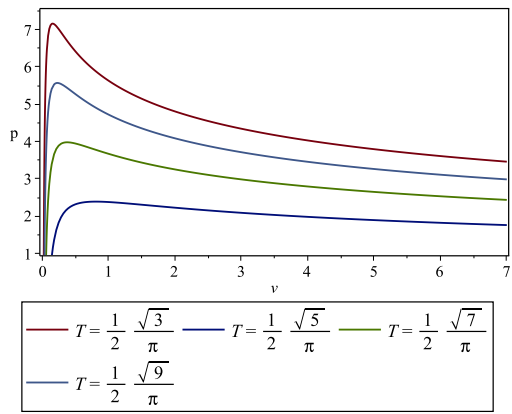}
\label{Fig2a}	

	\caption{Figure showing a P-V diagram for : The thermodynamic volume given by $V=\frac{4}{3}\pi r_{+}^3$ and pressure $P=\frac{3}{8\pi l^2}$  for SAdS blackholes  }
\end{figure}
\begin{figure}	
	\centering	
		\includegraphics[scale=0.35]{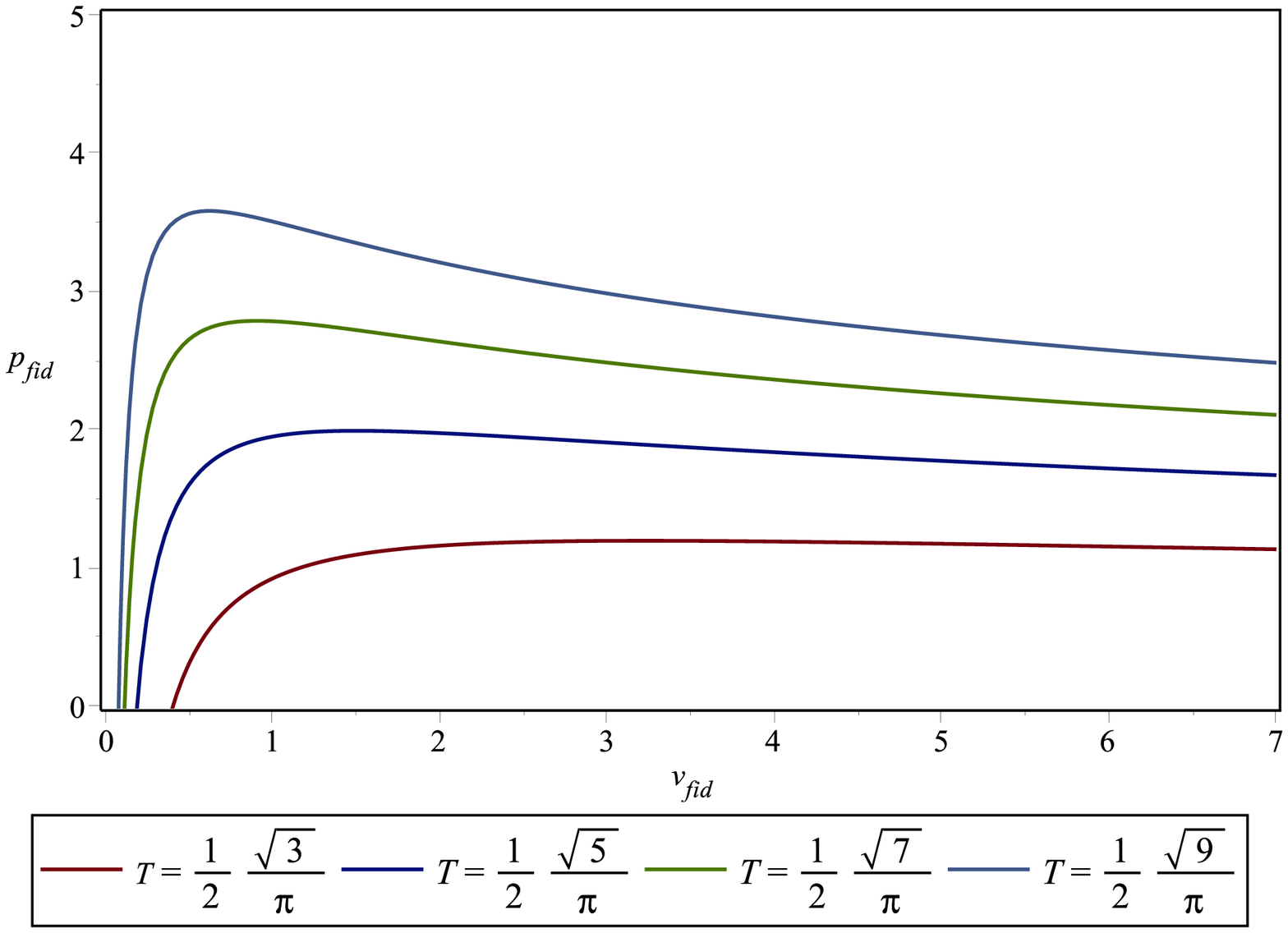}
\label{Fig2b}
	\caption{Figures showing a P-V diagram of fiedilty versus pressure. For various temperatures of SAdS blachoholes. Indicating that Fiedility does indeed represent thermodynamic volume.  }
\end{figure}
\begin{figure}
	\centering
		\includegraphics[scale=0.5]{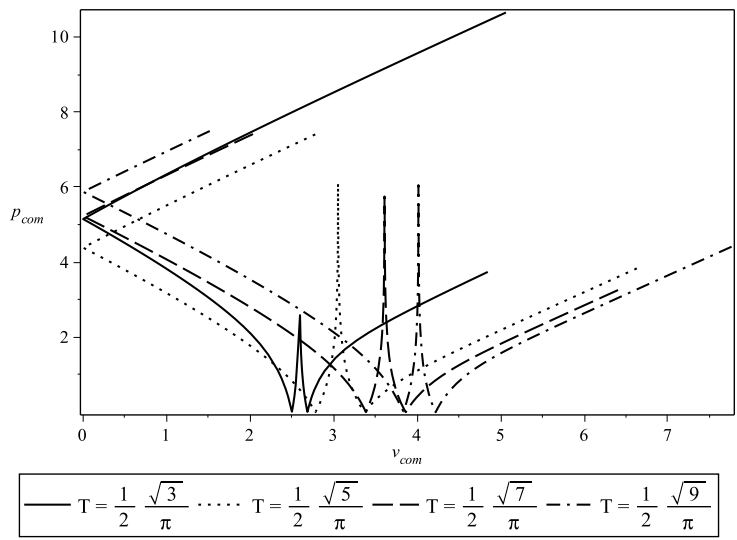}
		\caption{A P-V diagram between holographic complexity and pressure , showing a totally different behaviour than the thermodynamic P-V diagram. }
	\label{fig4}
\end{figure}
\\
Thus, we have calculated the holographic complexity and fidelity susceptibility from the associated minimal surface, and maximal volumes respectively and obtained a thermodynamic equation of states for SAdS black holes. We were also able to obtain the $P-V$ graphs for these quantaties.
, which are displayed in Figs.(1)-(6). 
It was observed that the $P-V$ relation for the fidelity susceptibility could be related to the 
$P-V$ relation for the themodynamic volume and pressure in extended phase space, where the   cosmological constant of the AdS space is viewed as a thermodynamic pressure \cite{ext1, ext2}. It may be noted that the dual theory to the Schwarzschild AdS has been studied \cite{b1}.
It would be possible to study the fidelity susceptibility of such a dual theory, 
and thus analyze the behavior of fidelity susceptibility. As this is the behavior 
obtained  from a well defined field theory, we expect it to be unitary. Now as the 
black hole thermodynamics in extended phase space also has the same behavior, 
we expect the black hole thermodynamics to also be a unitary process. 
 In the next section, we shall obtain the same calculation for a different deformation of AdS, with $ U(1)$ charged black holes. 

\section{Reissner-Nordstr\"{o}m Anti-de Sitter black holes (RNSAdS)}
The aim of this section is to extend the previous section to a  charged AdS black hole background. We will obtain fidelity susceptibility and holographic complexity and holographic equations of state.\\
The metric for RNSAdS is given by Eq.~(\ref{metric1}) with the function $f$ being
\be \label{ff}
f=1+{\frac {{r}^{2}}{{l}^{2}}}-{\frac {\epsilon{r_+^3}}{r{l}^{2}}}+{
\frac {\delta{r_+^4}}{{r}^{2}{l}^{2}}}\,,
\ee
where $\epsilon$ and $\delta$ are defined as
\be \label{epsilon}
\frac{2Ml^2}{r_+^3}=\epsilon \,, 
\qquad \frac{Q^2l^2}{r_+^4}=\delta\,.
\ee
Here, we also have that $|Q|<l/6$ and correspondingly $\delta<l^2/(6r_+^2)$. 
Let us start to compute extremal surfaces using the area functional. 
The area functional for a specific entangled region of the boundary in  RNSAdS is then given by Eq. (\ref{area1}) where $L(\theta)$ now is
\be
L(\theta)=r\sin\theta \sqrt{r^2+\frac{(\frac{dr}{d\theta})^{2}}{1+\frac{r^2}{l^2}-\frac{\epsilon r_+^3}{rl^2}+\frac{\delta r_+^4}{r^2l^2}}}\,.
\ee 
The Euler-Lagrange equation corresponding to the above $L(\theta)$ is written in the appendix (see \ref{EL}). We can solve this equation by expanding in series of $\theta$. Doing this, up to sixth order in $\theta$ we find (we set here the AdS radius $l=1$)
\begin{eqnarray} 
r \left( \theta \right)&=&\rho+\frac{1}{2}{\frac { \left(-\rho\epsilon{r_+
^3}+\delta{r_+^4}+{\rho}^{4}+{\rho}^{2} \right) {\theta}^{2
}}{\rho}}+\frac {1}{96\rho^2}\Big[ \big( 9\rho{
\epsilon}^{2}{r_+^6}-9\epsilon{h}^{7}\delta-45{\rho}^{4}\epsilon
\,{r_+^3}-29{\rho}^{2}\epsilon{r_+^3}\nonumber\\
&&+36{\rho}^{3}\delta
{r_+^4}+20\rho\delta{r_+^4}+36{\rho}^{7}+56{\rho}^{
5}+20{\rho}^{3} \big) {\theta}^{4}\Big]+\mathcal{O} \left( {\theta}^{6} \right)\,.\label{rtheta}
\end{eqnarray}
Now, by expanding the above expression in $\theta$ up to sixth  order, and then again expanding it in  $\delta$ up to second order, we find that the finite part of the entanglement entropy. This part is the difference between the pure background and the AdS deformation of the metric. Doing that, we find
\begin{eqnarray}
\Delta S&=&-\frac{1440\rho}{4G}\times[{\theta_0^2} ( 675\,{\theta_0^4}{r_+^3}\epsilon\,{
\rho}^{4}+540\,{\theta_0^2}{r_+^3}\epsilon\,{\rho}^{2}+
375\,{\theta_0^4}{r_+^7}\epsilon\,\delta+495\,{\theta_0^4}
{r_+^3}\epsilon\,{\rho}^{2}-450\,{\theta_0^4}{\rho}^{7}\nonumber\\
&&-
272\,{\theta_0^4}{\rho}^{3}-720\,{\theta_0^4}{\rho}
^{5}-540\,{\rho}^{5}{\theta_0^2}-480\,{\rho}^{3}{\theta_0^2}-420\,{\theta_0^4}\rho\,\delta\,{r_+^4}-600\,{\theta_0^4}{\rho}^{3}\delta\,{r_+^4}\nonumber\\
&&-720\,{\rho}^{3}
-540\,\rho\,{\theta_0^2}\delta\,{r_+^4} )]^{-1}\,.
\end{eqnarray}
The complete expressions for $L(\theta)$ and the integral related with the area are displayed in the appendix \ref{EL} (see Eqs.~(\ref{L}) and (\ref{Lint})).\\
Let us know compute the holographic complexity and fidelity susceptibility dual volumes for a  RN black hole. These quantities can be written as follows
\begin{eqnarray}
V_c&=&2\pi\int_0^{\theta_0}\sin\theta d\theta\displaystyle\int_{r_+}^{r(\theta)}\frac{r^2 dr}{\sqrt{f}}\,,\label{Vc}\\
V_{Fid}&=&2\pi\displaystyle\int_0^{2\pi}\sin\theta d\theta\int_{r_+}^{r_\infty}\frac{r^2 dr}{\sqrt{f}}\,.\label{Ffid}
\end{eqnarray}
Now, by expanding the integrand $r^{2}/\sqrt{f}$ in Taylor series up to linear terms  in $\epsilon$ and $\theta$, we obtain 
\begin{eqnarray}
\frac{r^2}{\sqrt{1+{\frac {{r}^{2}}{{l}^{2}}}-{\frac {\epsilon{r_+^3}}{r{l}^{2}}}+{
			\frac {\delta{r_+^4}}{{r}^{2}{l}^{2}}}}}&=&-\frac{1}{4}\frac {l r_+^3\left(-2r^2\epsilon l^2-2 r^4
\epsilon+2r_+\delta r l^2+2r_+\delta r^3+3r_+^4\delta
\epsilon\right)}{r\left(l^2+r^2\right)^{5/2}}\,.
\end{eqnarray}
Hence, the  volume corrsponding to the holographic complexity of the RN black hole becomes ($l=1$)
\begin{eqnarray}\label{Vcc}
V_c&=&-\int_{0}^{\theta_{0}}\sin\theta d\theta\Big(\frac{1}{2}{\frac {{r_+^3}\epsilon {r}^{2}}{ \left( 1+{r}^{2}
 \right) ^{3/2}}}+\frac{1}{2}{\frac {{r_+^3}\epsilon}{ \left( 1
+{r}^{2} \right) ^{3/2}}}+\frac{1}{2}{\frac {\delta {r_+^4}r}{\sqrt {1+{r}^{2}}}}+\frac{1}{4}{\frac {\epsilon{r_+^7}\delta}{\left( 1+{
r}^{2} \right) ^{3/2}}}\nonumber\\
&&+\frac{3}{4}{\frac {\epsilon{r_+^7}\delta}{
\sqrt {1+{r}^{2}}}}-\frac{3}{4}\epsilon{r_+^7}\delta\ln\left[ {
\frac {2+2\sqrt {1+{r}^{2}}}{r}}
 \right]\Big)\Bigl\vert_{r_+}^{r(\theta)}\,.
\end{eqnarray}
After computing this integral and expanding up to sixth  order in $\theta$, we find expression (\ref{V1}). After some mathematical steps (explained in appendix \ref{EL}), by taking asymptotic expansion in $\rho$ we find the following compacted expression  
\begin{eqnarray}
V_c&=&-\frac{1}{48}{\frac { \left( -6\,{\theta_0^2}\epsilon\,b{r_+^2}-6\,{
\theta_0^2}\epsilon\,b \right) {\theta_0^2}{r_+^3}\pi \,
\rho}{ b^3}}\nonumber\\
&&-\frac{1}{48b^3}\times \Big[ -3\,{\theta_0^2}{r_+^6}\delta\,b\epsilon\,\ln  \left( {\frac {1+b}{r_+}} \right) 
+2\,{\theta_0^2}{r_+^4}\delta+4\,{\theta_0^2}{r_+^3}\delta
\,b-48\,{r_+^4}\delta\,\epsilon-24\,{r_+^4}\delta\nonumber\\
&&+2\,{\theta_0^2
}\epsilon\,{r_+^2}+4\,{\theta_0^2}r_+\delta\,b+24\,r_+\delta\,b-3\,{
\theta_0^2}{r_+^4}\delta\,b\epsilon\,\ln  \left( {\frac {1+b}{r_+}
} \right) +2\,{\theta_0^2}{r_+^2}\delta+4\,{\theta_0^2}{r_+^4}\delta\,\epsilon\nonumber\\
&&-24\,\epsilon+2\,{\theta_0^2}\epsilon-24\,
\epsilon\,{r_+^2}+24\,{r_+^3}\delta\,b-36\,{r_+^6}\delta\,\epsilon+36
\,{r_+^4}\delta\,\epsilon\,b\ln  \left( {\frac {1+b}{r_+}} \right) -24\,
{r_+^2}\delta \nonumber\\
&&+36\,{r_+^6}\delta\,\epsilon\,b\ln  \left( {\frac {1+b}{r_+
	}} \right)+3\,{\theta_0^2}{r_+^6}\delta\,\epsilon \Big] {
	\theta_0^2}{r_+^3}\pi\,,
\end{eqnarray}
where $a=\sqrt{1+\rho^2}$ and $b=\sqrt{1+r_+^2}$.\\
The mass, temperature and complexity pressure of the RN black hole are defined as follows
\begin{align}
M&=\frac{r_+}{2}\left( 1+{\frac {{r_+^2}}{{l}^{2}}}+{\frac {{Q}^{2}}{{r_+^2}}
} \right)\,,\\
T&=\frac{1}{4\pi}\left(\frac{3r_+^4+r_+^2-Q^2}{r_+^3}\right)\label{TT}\,,\\
P&=-\frac{\partial M}{\partial V_{Fid}}=-\frac{\frac{\partial M}{\partial r_{+}}}{\frac{\partial V_{Fid}}{\partial r_{+}}}\,.
\end{align}
The explicit expression for the complexity pressure is very long for the space-time studied. The complete and expanded expressions are displayed in the appendix \ref{EL}. Now, we need to express the complexity pressure and volume in terms of the temperature. In order to do that, we need to solve (\ref{TT}) for $r_+$. Thus, we need to solve the following equation
\be \label{eq}
r_+^4-\frac{4\pi T}{3}r_+^3+\frac{r_+^2}{3}-\frac{Q^2}{3}=0\,.
\ee
The roots of this equation are given by
\bea
x_{{1,2}}&=&-\frac{\tilde{b}}{4\tilde{a}}-S\pm\frac{1}{2}\sqrt {-4{S}^{2}-2p+{\frac {q}{S}}}\,, \nn\\
\qquad x_{{3,4}}&=&-\frac{\tilde{b}}{4\tilde{a}}+S\pm\frac{1}{2}\sqrt {-4{S}^{2}-2p+{\frac {q}{S}}}\,,
\eea
where $p$, $q$ and $S$ are defined by
\bea
p&=&\frac{8\tilde{a}\tilde{c}-3\tilde{b}^2}{8\tilde{a}^2}\,,\qquad q=\frac{\tilde{b}^3-4\tilde{a}\tilde{b}+8\tilde{a}^2\tilde{d}}{8\tilde{a}^3}\,,\nn\\
S&=&\frac{1}{6}\sqrt {-6+3G+3{\frac {\Delta_{{0}}}{G}}}\,, 
\qquad G=\sqrt [3]{\frac{1}{2}\Delta_{{1}}+\frac{1}{2}\sqrt {{\Delta_1^2}-4{
\Delta_0^3}}}\,, \nn\\
\Delta_{{0}}&=&\tilde{c}^2-3\tilde{b}\tilde{d}+12\tilde{a}\tilde{e}\,, \qquad \Delta_1=2\tilde{c}^3-9\tilde{b}\tilde{c}\tilde{d}+27\tilde{b}^2\tilde{e}+27\tilde{a}\tilde{d}^2-72\tilde{a}\tilde{c}\tilde{e}\,,
\eea and $\Delta$ determined as
\be
\Delta_1^2-4\Delta_0^3=-27\Delta\,
\ee 
is a determinant of the fourth order polynomial. If $\Delta>0$, then all four roots of the equation are either real or complex.
From (\ref{eq}), we have that $\tilde{a}=1$, $\tilde{b}=-(4\pi T)/3$, $\tilde{c}=1/3$, $\tilde{d}=0$ and $\tilde{e}=-Q^2/3$. Therefore, for our case the roots are given by
\begin{eqnarray}\label{r+12}
r_+^{1,2}&=&\frac{1}{3}\pi \,T-\frac{1}{6}\sqrt {-6+3\,\sqrt [3]{k}+3\,{\frac {\frac{1}{9}-4\,{Q}^{2}}
{\sqrt [3]{k}}}}\nonumber\\
&&\pm\frac{1}{6}\sqrt {-3\,\sqrt [3]{k}-3\,{\frac {\frac{1}{9}-4\,{Q}^{2
}}{\sqrt [3]{k}}}+12\,{\pi }^{2}{T}^{2}+ {\frac {54\, \left( -{\frac {8}{27}}\,
{\pi }^{3}{T}^{3}+\frac{2}{3}\pi \,T \right)}{\sqrt {-6+3\,\sqrt [
3]{k}+3\,{\frac {1/9-4\,{Q}^{2}}{\sqrt [3]{k}}}}}}}\,,
\end{eqnarray}

\begin{eqnarray}
r_+^{3,4}&=&\frac{1}{3}\pi \,T-\frac{1}{6}\sqrt {-6+3\,\sqrt [3]{k}+3\,{\frac {\frac{1}{9}-4\,{Q}^{2}}
{\sqrt [3]{k}}}}\nonumber\\
&&\pm\frac{1}{6}\sqrt {-3\,\sqrt [3]{k}-3\,{\frac {\frac{1}{9}-4\,{Q}^{2
}}{\sqrt [3]{k}}}+12\,{\pi }^{2}{T}^{2}+ {\frac {54\, \left( -{\frac {8}{27}}\,
{\pi }^{3}{T}^{3}+\frac{2}{3}\pi \,T \right)}{\sqrt {-6+3\,\sqrt [
3]{k}+3\,{\frac {1/9-4\,{Q}^{2}}{\sqrt [3]{k}}}}}}}\,,
\end{eqnarray} where
\be
k=\frac{1}{27}-8\,{\pi }^{2}{T}^{2}{Q}^{2}+4\,{Q}^{2}+\frac{1}{2}\sqrt { \left( {
\frac {2}{27}}-16\,{\pi }^{2}{T}^{2}{Q}^{2}+8\,{Q}^{2} \right) ^{2}-4
\, \left(\frac{1}{9}-4\,{Q}^{2} \right) ^{3}}.
\ee
Since $k$ must be real, the inequality
\be
\left( {\frac {2}{27}}-16\,{\pi }^{2}{T}^{2}{Q}^{2}+8\,{Q}^{2}
\right) ^{2}-4\, \left( \frac{1}{9}-4\,{Q}^{2} \right) ^{3}\geq0
\ee
must hold. Equivalently, the above inequality can be expressed as 
\be
T\leq\frac{1}{4Q\pi}\sqrt {- \left( 2\, \left(\frac{1}{9}-4\,{Q}^{2} \right) ^{3/2}-{\frac {
2}{27}}-8\,{Q}^{2} \right)}\,.
\ee
Now, by taking series in $Q$ up to order 6 in the above equation, we find
\be
T\leq\frac{1}{2}\,{\frac {\sqrt {3}}{\pi }}-\frac{3}{4}\,{\frac {\sqrt {3}{Q}^{2}}{\pi }}
-{\frac {81}{16}}\,{\frac {\sqrt {3}{Q}^{4}}{\pi }}+\mathcal{O} \left( {Q}^{6}\right),
\ee or
\be
T\leq0.27567- 0.41350\,{Q}^{2}- 2.7912\,{Q}^{4}+\mathcal{O} \left( {Q}^{6} \right).
\ee
From this expression (see Fig. (7)), we can note that the temperature will be maximum when $Q=0$. 
\begin{figure}[ht]
	\centering
	\includegraphics[width=80mm]{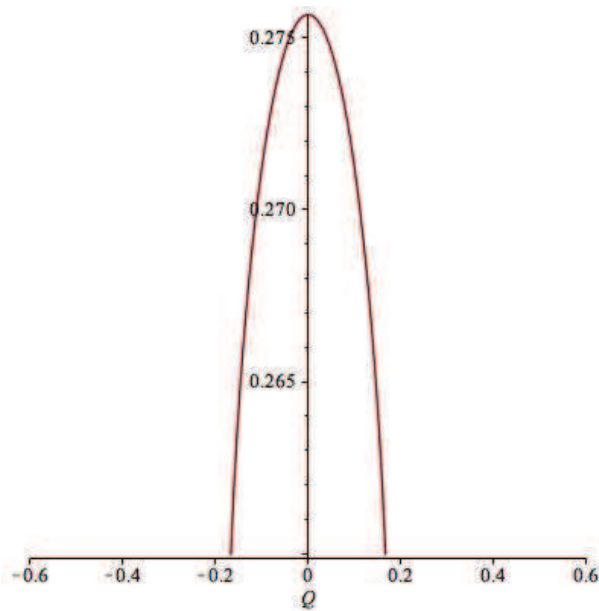}
	\caption{$T$ as a function of  $Q$ . This graph is confined with the condition $|Q|<\frac{l}{6}$.}
	\label{constlat}
\end{figure}
Let us now find the horizon of RNSAdS black hole. To do that, we need to rewrite (\ref{ff}) using (\ref{epsilon}), which gives us
\be
f(r)=1+\frac{r^2}{l^2}-\frac{2M}{r}+\frac{Q^2}{r^2}\,.
\ee 
Now, the horizon condition $f(r)=0$ can be written as
\be
\xi^4+\xi^2-\frac{2M}{l}\xi+\left(\frac{Q}{l}\right)^2=0\,,
\ee
where $\xi=r/l$. Here our aim is to find the largest root of the above equation which will correspond to the outer horizon of the RN charged black hole. The largest root is given by
\begin{eqnarray}
\xi_3=&\frac{1}{6}\left(-6+3k+3\frac {1+12Q^2}{
k}\right)^{1/2}+\frac{1}{6}\sqrt {-12-3m-
\frac {3\left(1+12{Q}^{2}\right)}{m}-{\frac {108M}{\sqrt {
-6+3m+3{\frac {1+12{Q}^{2}}{m}}}}}}\,,\nonumber\\
\end{eqnarray}
where $m=\left(1+54\,{M}^{2}+6\,\sqrt {3\,{M}^{2}+81\,{M}^{4}-{Q}^{2}-12\,{
Q}^{4}-48\,{Q}^{6}}\right)^{1/3}$. 
This root can be seen in Figs.~(8)-(10).
\begin{figure}[ht!]
	\centering
\includegraphics[scale=0.5]{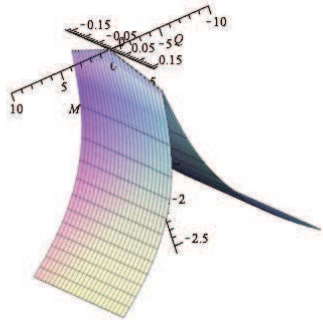}
\label{Fig3}
\caption{Location of the roots $\xi_{1}$ for the RN black hole as a function of $Q,M$.}
\end{figure}
\begin{figure}[ht!]
	\centering
\includegraphics[scale=0.5]{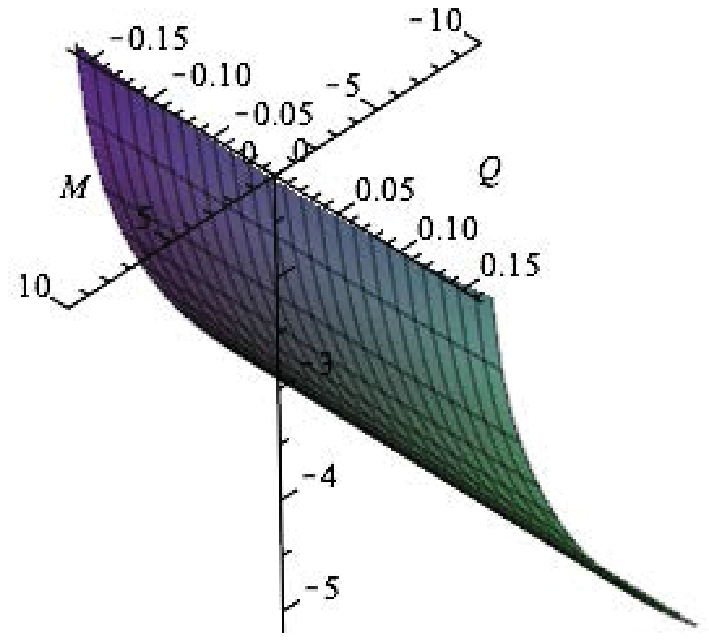}
\label{Fig3a}
\caption{Location of the roots $\xi_{2}$ for the RN black hole as a function of $Q,M$. }
\end{figure}
\begin{figure}[ht!]
	\centering
\includegraphics[scale=0.55]{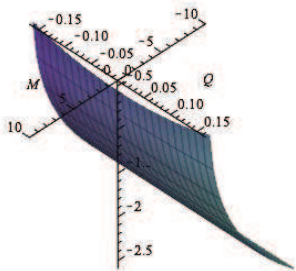}
\caption{Location of the roots $\xi_{3}$ for the RN black hole as functions of $Q,M$. Showing that $\xi_3$ is the largest root.}
\label{Fig3aa}
\end{figure}
\begin{figure}
	\centering
		\includegraphics[scale=0.35]{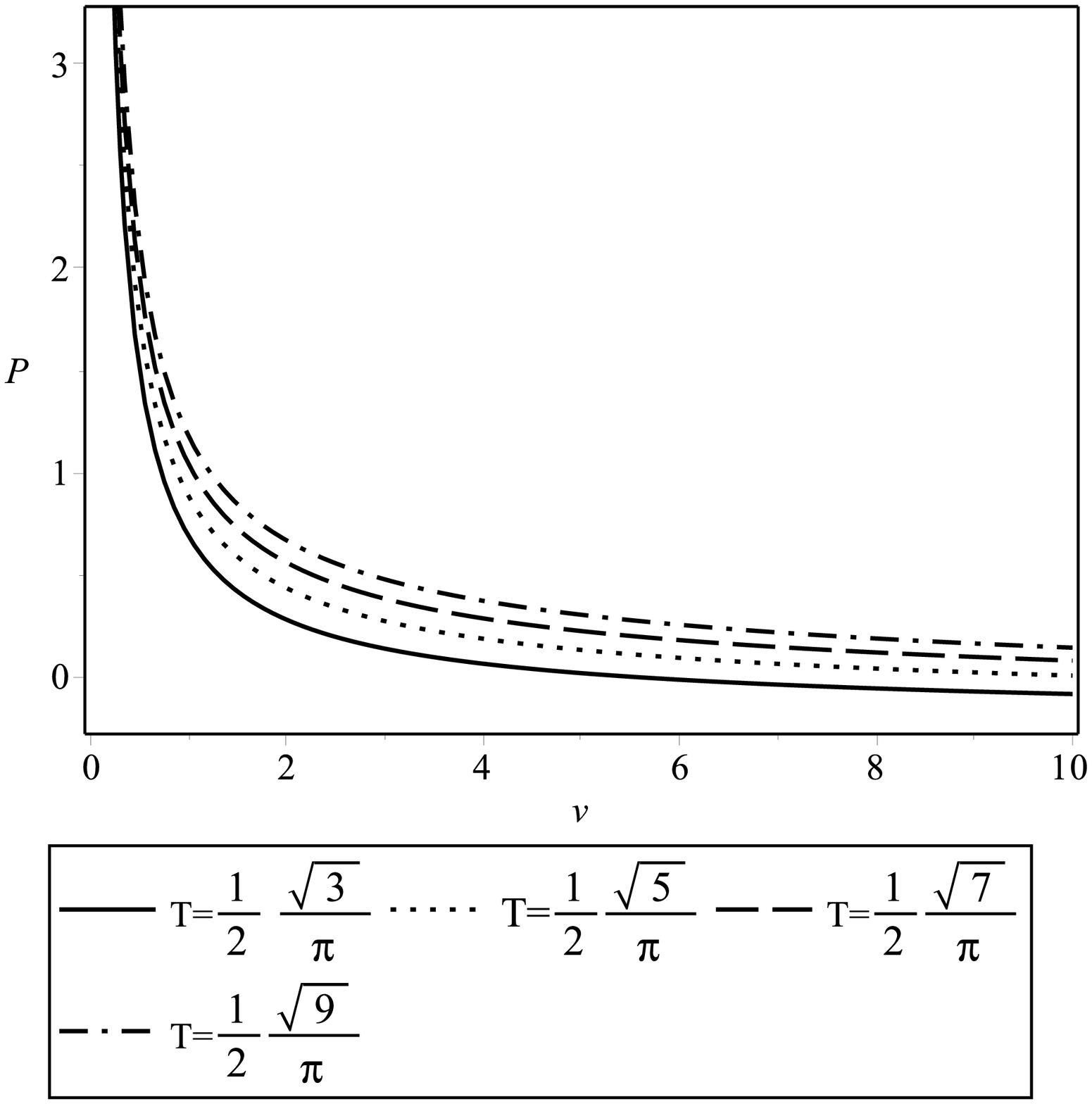}
			\caption{Figure showing a P-V diagram for  The thermodynamic volume $V=\frac{4}{3}\pi r_{+}^3$ and pressure $P=\frac{3}{8\pi l^2}$ for RNSAdS blackholes.}
			\label{Fig4}
		\end{figure}
		\begin{figure}
			\centering
	\includegraphics[scale=0.35]{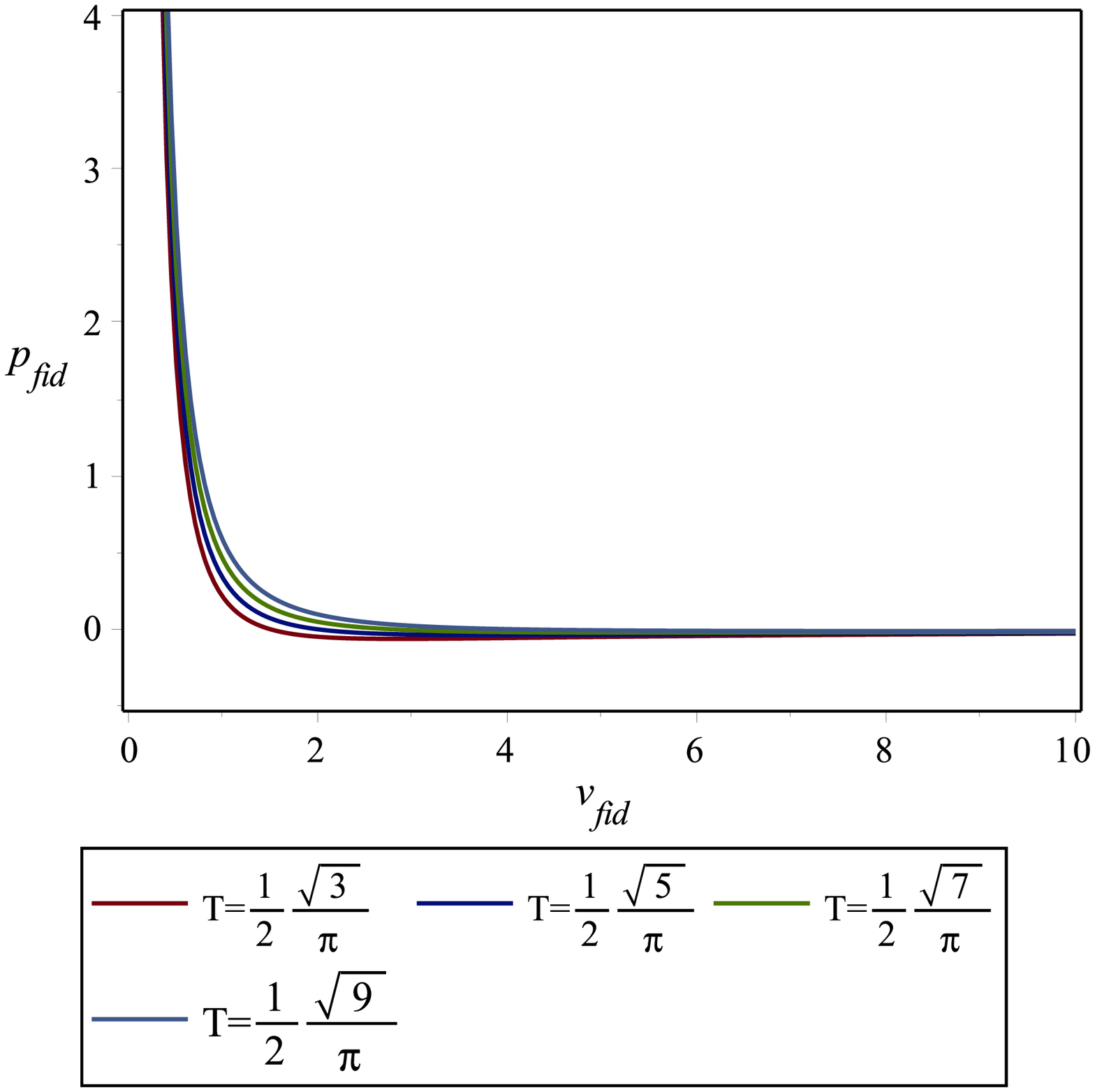}
	\caption{Figure showing a P-V diagrams for fiedilty versus pressure . For various temperatures of RNSAdS blachoholes. Indicating that Fidelity does indeed represent thermodynamic volume.  }
	\label{Fig4a}
\end{figure}
\begin{figure}
\centering
		\includegraphics[scale=0.5]{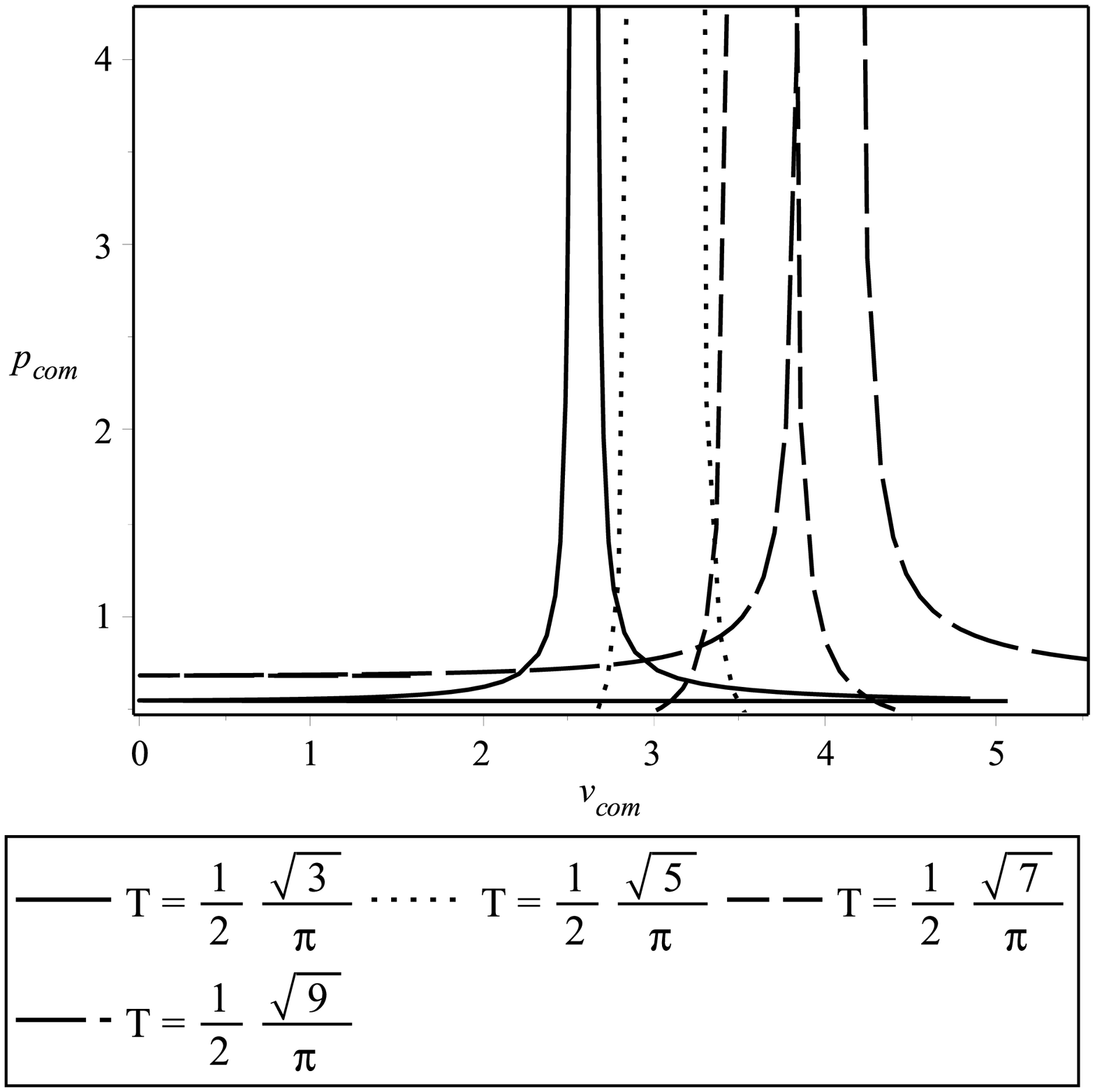}
		\caption{A P-V diagram between holographic complexity and pressure , showing a totally different behaviour than the thermodynamic P-V diagram. }
		\label{Fig44}
\end{figure}
From Fig.~\ref{Fig4} and ~\ref{Fig4a}, 
We again observe that he $P-V$ behavior  of  fidelity susceptibility is similar to the $P-V$ relation  of  thermodynamic volume and pressure . However
, in Fig.~\ref{Fig44}, 
we observe that the $P-V$ relation obtained from  holographic complexity is very different.
The dual to a Reissner-Nordstr\"om-AdS black hole has been studied \cite{rn,rnrn}, 
and it is possible to obtain the fidelity susceptibility of  this 
dual theory. This should correspond to the fidelity susceptibility calculated 
from the bulk. Now as the behavior of fidelity susceptibility from the 
dual theory cannot break unitarity, the behavior of fidelity susceptibility in the 
bulk cannot also break unitarity. However, as the behavior of fidelity susceptibility 
resembles the behavior of thermodynamics in extended phase space, it can be argued that 
the thermodynamics of a  Reissner-Nordstr\"om-AdS black hole  is represented 
by a unitarity process. This can be used as a proposal to resolve the black hole 
information paradox in a  Reissner-Nordstr\"om-AdS black hole.

\section{SAdS for any dimensions}
In this section we will extend our previous result to higher dimensional AdS space times when there is no electric charge. The line-element of a Schwarzschild AdS spacetime for any dimension $n$ can be written as follows 
\be
ds^2=-f(r)dt^2+\frac{dr^2}{f(r)}+r^2d\Omega_{\left(n-2\right)},
\ee
where $d\Omega_{n-2}$ denotes the metric of a $S_{n-2}$ sphere defined as
\begin{eqnarray}
d\Omega_{n-2}=d\theta_{1}^2+\frac{1}{k}\sin^2(\sqrt{k}\theta_{1})d\theta_{2}^2+\frac{1}{k}\prod_{i=1}^{n-3}\sin^2(\sqrt{k} \theta_{i})d\theta_{n-2}\,,
\end{eqnarray}
where $k=\{-1,0,1\}$ and the function $f(r)$ is
\begin{eqnarray}
f(r)=1-\frac{2M}{r^{n-3}}+\frac{r^2}{l^2}\,.
\end{eqnarray}
Clearly, if we set $n=4$ we recover the case studied in Sec.~\ref{Sads}. Now, at the horizon $r_{H}$, the function $f(r)=0$ so that the mass of the black hole and its horizons satisfy the following equation
\be
M=r_H^{n-3}\left(1+\frac{r_H^2}{l^2}\right)\,.
\ee
Let us now use the same approach that we used before to compute the area of the minimal surface $\gamma_{A}$. By taking a time slice at $t=0$ in the above line-element, we obtain
\be
ds^2|_{t=0}=\left(\frac{\left(\frac{dr}{d\theta}\right)^2}{f(r)}+r^2\right)dx^2+r^2(x)\underbrace{\sum_{a=2}^{n-2}\left(dx^a\right)^2}_{n-3}\,
\ee
where we have parametrised the surface as $r=r(\theta)$ and we have assumed that the $x^{a}$ coordinate lies between $-L/2\leq x^a\leq L/2$ where $L$ is the total entangled length of the subsystem on boundary. Hence, the area of the minimal surface can be expressed as follow
\be
\mbox{Area}\equiv A=L^{n-3}\int r^{n-3}\sqrt{r^2+\frac{r'^2}{f(r)}}dx\,.\label{area2}
\ee
Note that in the above equation, the Lagrangian density $L=L(r,r')$ does not have the coordinate $x$, i.e.  
 $\frac{\partial L}{\partial x}=0$, so that this term can be written outside the integral. Explicitly, this Lagrangian has a conserved charge
. If we think on $x$ as time coordinate in dynamical system appproach, it needs to satisfy the following constraint,
\be
r'\frac{\partial L}{\partial r'}-L=C\equiv\mbox{const}\,\label{energy},
\ee
and therefore we have
\be
\frac{\partial L}{\partial r'}=r^{n-3}\frac{\frac{r'}{f(r)}}{\sqrt{r^2+\frac{r'^2}{f(r)}}}.
\ee
The term in the integrand  in Eq. (\ref{area2}) can be simplified as
\be
L^{n-3}\{ r^{n-3}\frac{\frac{r'^2}{f(r)}}{\sqrt{r^2+\frac{r'^2}{f(r)}}}-r^{n-3}\sqrt{r^2+\frac{r'^2}{f(r)}} \}=L^{n-3}\{r^{n-3}\}\left(-\frac{r^2}{\sqrt{r^2+\frac{r'^2}{f(r)}}}\right)=C.
\ee
The appropriate boundary conditions are given at the turning point $r^{*}$ as $r'|_{r=r^*}=0$ and using this boundary condition in (\ref{energy}), we obtain
\be
 C=L^{n-3}(r_*)^{n-3}(-r_*)=-L^{n-3}(r_*)^{n-2}.
\ee
Using this expression we are able to more simplify the integrand in  Eq. (\ref{area2}) as follows
\be
A=2L^{n-3}\int_0^{r^*}r^{n-3}xr\left(\frac{r}{r^*}\right)^{n-2}dx=2x\frac{L^{n-3}}{r^{*n-2}}\int_0^{r^*}r^{2n-4}\frac{dx}{dr}dr=2x\frac{L^{n-3}}{r^{*n-2}}\int_0^{r^*}(r(x))^{2n-4}dx
\ee
The corresponding maximal volume is given by the following expression,
\be
V_{max}=\int\frac{dr}{\sqrt{f(r)}}r^{n-1}\int dx L^{n-3}=L^{n-3}\int_0^{r^*}\frac{dr}{\sqrt{f(r)}}r^{n-1}x(r).
\ee
Now we consider  case in which $n=5$. Five dimensional black objects widely studied in literature because they could have different topologies than case in $n=4$ \cite{Emparan:2008eg}. 
\begin{figure}
\centering
		\includegraphics[scale=0.35]{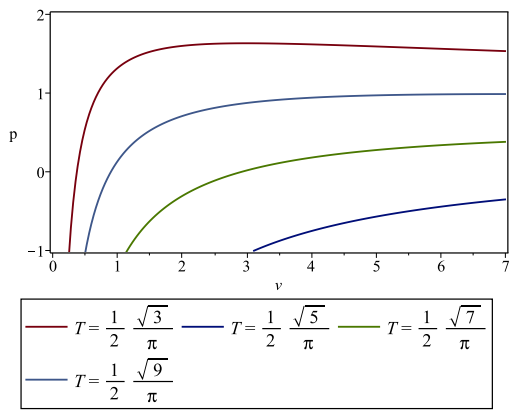}
			\caption{Figure showing P-V diagrams for the thermodynamic volume and pressure for  five dimensional SAdS blackholes }
			\label{fig9a}	
		\end{figure}
\begin{figure}
	\centering
		\includegraphics[scale=0.35]{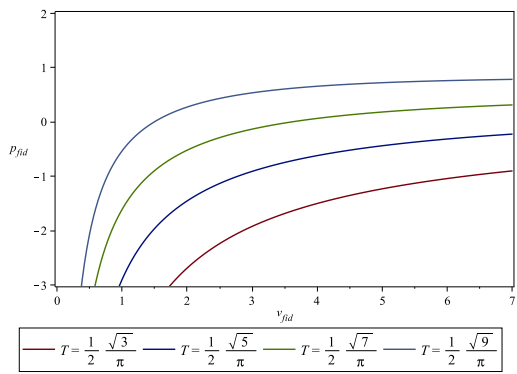}
\label{fig9b}	
	\caption{A figure for  five dimensional SAdS blackholes showing the behavior of fiedilty susceptibility versus pressure. For various temperatures of  this higher dimensional  black holes. Indicating that the association of fiedility susceptibility to thermodynamic volume is universal to any dimension. }
\end{figure}
As the previous sections, we are able to define the fidelity pressure using $V_{max}$. The complete expression is written in the appendix \ref{sAdsanyd}. From Figs.~(13)-(15), 
It may be noted that here again we observe that the behavior of fidelity susceptibility is similar to the  thermodynamics volume. However, the behavior of holographic complexity is very different from the behavior of both the fidelity susceptibility and the thermodynamic volume.  This indicates that the behavior is not a property of the specific metric, but seems to be a universal behavior of fidelity susceptibility, holographic complexity, and thermodynamic volume. 
 The dual to a higher dimensional Schwarzschild AdS black hole has also been constructed
\cite{12hd, hd12}. It is possible to analyze the behavior of  
 fidelity susceptibility of  this 
dual theory, and it this would be described 
by a unitarity process. As the behavior of a fidelity susceptibility of a 
higher dimensional Schwarzschild AdS black hole  is similar to  the behavior of its  thermodynamics 
in extended phase space, 
the thermodynamics of a  Schwarzschild AdS black hole is expected to be described by 
a unitarity process. Furthermore, as this universal behavior seems, and not a property of a specific metric, 
this can  be used as a proposal for the black hole information 
paradox. 

\section{Conclusions}
In this paper, we studied the connection between the information dual to different volumes 
in the bulk of a deformed AdS spacetime and the thermodynamic volume in extended phase space. 
In the extended phase space, the cosmological constant can be related to the thermodynamic pressure, 
and a conjugate thermodynamic volume for this pressure can be defined. 
Furthermore, the information dual to different volumes in the bulk AdS space was measured 
by the fidelity susceptibility and the holographic complexity. 
As these quantities diverged, we used a regularized definition for these quantities. 
We regularized them by subtracting the contribution from the background AdS spacetime 
from the deformation of the AdS spacetime. It was possible to use this regularized fidelity 
susceptibility and regularized holographic complexity  to define pressure for these 
quantities. Thus, we analyzed the $P-V$ equation for these quantities, and compared it 
with the $P-V$ form for the thermodynamic volume 
and thermodynamic pressure in the extended phase space. 
It was observed that the regularized fidelity susceptibility had the same behavior as the thermodynamic 
volume. The regularized holographic complexity had a very different behavior. We observed 
this phenomenon for different black holes. This indicates that this is a universal  behavior of the 
fidelity susceptibility, holographic complexity, and thermodynamic volume, and not a property 
of the specific geometry. 
It may be noted that fidelity susceptibility of the 
bulk has a well defined boundary dual. Thus, as the fidelity 
susceptibility and the thermodynamic volume of the black holes show 
similar behavior, it is expected that both of them represent the same 
physical quantity. Now it is known that the evolution of the fidelity 
susceptibility can be understood in the dual picture from the evolution 
of a conformal field theory. As the evolution of the conformal field theory 
is a unitary process, it is expected that the evolution of the fidelity 
susceptibility in the bulk will also be a unitary process. Now as the fidelity 
susceptibility can be related to the black hole thermodynamics, it can be argued 
that the black hole thermodynamics in the extended phase space would be dual 
to a unitary process. This might help resolve black hole information paradox. 

It would be interesting to generalize this analysis to time-dependent geometries. 
The time-dependent holographic complexity for such time-dependent geometries has 
been recently studied \cite{timedepn}. It is possible to study the fidelity susceptibility
of such time-dependent geometries. It would be interesting to analyze the   behavior of 
$P-V$ diagrams for the 
holographic complexity and the 
fidelity susceptibility for such time-dependent geometries.
It would also be interesting to analyze the thermodynamics 
of black holes in such a time-dependent geometry, and compare the 
thermodynamics of black hole in extended phase space to fidelity susceptibility and 
holographic complexity. It is expected that again the fidelity susceptibility and 
the thermodynamics in extended phase space will have similar behavior, and the holographic 
complexity will have a very different behavior. 

 The black hole thermodynamics has been studied in massive gravity \cite{a2m1}. 
 It would be interesting to obtain the fidelity susceptibility and the holographic 
 complexity for black hole geometries in massive gravity. It is again extended 
 that these will diverge, but can be regularized by subtracting the contribution 
of background AdS spacetime from the deformation of the AdS spacetime. It would be interesting 
to analyze the behavior of the $P-V$ diagrams for the fidelity susceptibility and the holographic 
 complexity in massive gravity. This can be compared to the thermodynamics in extended phase space.
 It is important to understand the behavior of such 
   $P-V$ diagrams for massive gravity, to understand if the similar behavior of 
   fidelity susceptibility and thermodynamics in extended phase space is restricted 
   to the Einstein gravity,  if it also exists in massive gravity. 
   Complexity was extensively studied for many interesting physical systems like in one-dimensional holographic superconductors~\cite{Zangeneh:2017tub}, charged black holes~\cite{Couch:2016exn,Okcu:2017hnp}, black holes in $f(R)$ gravity \cite{Guo:2017rul}, singular surfaces~\cite{Bakhshaei:2017qud}, AdS black holes~\cite{Mo:2017oqj,Bhattacharya:2017hfj} and massive gravity~\cite{Pan:2016ecg}. It would be interesting to extend our work to there systems.
 
\section*{Acknowledgments}
The authors thank Sebastian Bahamonde for contributions in an early draft of this work.

\appendix 
\section{SAdS}\label{appendixA}
The expressions of volume and pressure are
\begin{eqnarray}
V=\frac{1}{2}\left( -{\frac { \left( \frac{2}{3}\pi \,T-\frac{1}{3}\sqrt {4\,{\pi }^{2}{T
			}^{2}-3} \right) ^{3}}{\sqrt {1+{\rho}^{2}}}}+{\frac { \left( \frac{2}{3}
		\pi \,T-\frac{1}{3}\sqrt {4\,{\pi }^{2}{T}^{2}-3} \right) ^{3}}{\sqrt {
			\left( \frac{2}{3}\pi \,T-\frac{1}{3}\sqrt {4\,{\pi }^{2}{T}^{2}-3} \right) ^{2}+
			1}}} \right) {\theta_0^2},
\end{eqnarray}
\begin{eqnarray}
P&=&-{\frac{a\left(1+3\left(\frac{2}{3}\pi T-\frac{1}{3}\sqrt{4{\pi }^{2}{T
			}^{2}-3}\right)^{2}\right)\left(\left(\frac{2}{3}\pi T-\frac{1}{3}\sqrt{4\pi^{2}T^2-3}\right)^{2}+1\right)^
		{3/2}}{\left(\frac{2}{3}\pi T-\frac{1}{3}\sqrt{4{\pi}^{2}T^2-3}
		\right)^{4}}}\nonumber\\
&&\times\frac{1}{\left(-3\sqrt{\left(\frac{2}{3}\pi T-\frac{1}{3}\sqrt {4{
				\pi}^2{T}^{2}-3} \right)^{2}+1}-\frac{3\sqrt { \left(\frac{2}{3}\pi T-\frac{1}{3}\sqrt {4{\pi }^{2}{T}^{2}-3} \right)^{2}+1}}{\left( \frac{2}{3}\pi T-\frac{1}{3}\sqrt {4{\pi }^{2}{T}^{2}-3} \right) ^{2}}+2a+\frac{3a}{\left(\frac{2}{3}\pi T-\frac{1}{3}\sqrt {4{\pi }^{2}{T}^{2}-3} \right)^{2}} \right) {\theta_0^2}},\nonumber\\ \nonumber \\
\end{eqnarray} 
where $a=\sqrt{1+\rho^2}$.
\section{RNSAdS}
E-L equation:
\begin{eqnarray}\label{EL}
{\frac {d^{2}r}{d{\theta}^{2}}}&=&\frac{1}{-2r^{2}\sin \left( \theta
	\right) {l}^{2}\epsilon{r_+^3}+2r\sin
	\left( \theta \right) {l}^{2}\delta{r_+^4}+2 r^{5}\sin \left( \theta \right) {l}^{2}+2r^{3}\sin \left( \theta
	\right) {l}^{4}}\nonumber\\
&&\times[4\sin \left( 
\theta \right) r^{8}+8\sin
\left( \theta \right) {l}^{2} r 
^{6}-8r^{5}\sin
\left( \theta \right) \epsilon{r_+^3}-2r^{5}\cos \left( \theta \right)  \left( {\frac {dr}{d
		\theta}}\right) {l}^{2}+4r^{4}\sin \left( \theta \right) {l}^{4}\nonumber\\
&&+8
r^{4}\sin \left( \theta
\right)  \left( {\frac {dr}{d\theta}} \right) 
^{2}{l}^{2}+8r^{4}\sin
\left( \theta \right) \delta{r_+^4}-2 r^{3}\cos \left( \theta \right)  \left( {\frac {dr}{d
		\theta}}\right) {l}^{4}-8r^{3}\sin \left( \theta \right) {l}^{2}\epsilon
{h}^{3}\nonumber\\
&&+2r^{2}\cos \left( 
\theta \right)  \left( {\frac {dr}{d\theta}}
\right) {l}^{2}\epsilon{r_+^3}+4r^{2}\sin \left( \theta \right) {\epsilon}^{2}{r_+^6}+6
r^{2}\sin \left( \theta
\right)  \left( {\frac {dr}{d\theta}} \right) 
^{2}{l}^{4}+8r^{2}\sin
\left( \theta \right) {l}^{2}\delta{r_+^4}\nonumber\\
&&-5r\sin \left( \theta \right)  \left( {\frac {dr}{d\theta}}\right) ^{2}{l}^{2}\epsilon{r_+^3}-2r
\left( \theta \right) \cos \left( \theta \right)  \left( {\frac {dr}{d
		\theta}}\right) {l}^{2}\delta{r_+^4}-8r\sin \left( \theta \right) \epsilon{r_+^7}
\delta \nonumber\\
&&-2r\cos \left( \theta \right)  \left( {
	\frac {dr}{d\theta}}\right)^{3}{l}^{4}+4
\sin \left( \theta \right) {\delta}^{2}{r_+^8}+4\sin \left( \theta
\right)  \left( {\frac {dr}{d\theta}}\right) 
^{2}{l}^{2}\delta{r_+^4}].
\end{eqnarray}
The functional $L(\theta)$ with the function $r(\theta)$ given by (\ref{rtheta}) is
\begin{eqnarray}
L&=&-\frac {3}{32\rho^2}\nonumber\\
&&\times[-4\theta^4\rho^7-\frac {56}{9}\theta^2
\left(\frac{6}{7}+\theta^2 \right) \rho^5+5\theta^4\rho^4
\epsilon r_+^3+ \left(  \left( -4\delta r_+^4-\frac {20}{9}
\right) \theta^4-\frac{16}{3}\theta^2-\frac {32}{3} \right) \rho
^3\nonumber\\
&&+{\frac {29}{9}}{r_+^3} \left( {\theta}^{2}+{\frac {48}{29}}
\right) {\theta}^{2}\epsilon{\rho}^{2}-{r_+^4}{\theta}^{2} \left( 
\left( {\epsilon}^{2}{r_+^2}+{\frac {20}{9}}\delta \right) {\theta}
^{2}+\frac{16}{3}\delta \right) \rho+{\theta}^{4}\epsilon{r_+^7}\delta
] \nonumber\\
&&\times\sin \left( \theta \right) \sqrt { \left( \rho+c-d \right) ^{
		2}+\frac{(2c-4d)^2}{\left( 1+ \left( \rho+c-d \right) ^{2}-{\frac {
				\epsilon{r_+^3}}{\rho+c-d}}+{\frac {\delta r_+^4}{ \left( \rho+c-
				d \right) ^2}} \right)\theta^2}}
\end{eqnarray}
where
\begin{eqnarray}
	c&=&\frac{1}{2}{\frac { \left( -\rho\,\epsilon\,{r_+^3}+\delta\,{r_+^4}+{\rho}^{
				4}+{\rho}^{2} \right) {\theta}^{2}}{\rho}},
	\\
	d&=&\frac {1}{96\rho^2}\Big[ \big( -9\rho\epsilon^2r_+^6+9
			\epsilon r_+^7\delta+45\rho^4\epsilon r_+^3+29\rho^2
			\epsilon r_+^3\nonumber\\
			&&-36\rho^3\delta r_+^4-20\rho\delta r_+^
			4-36\rho^7-56\rho^5-20\rho^3 \big) \theta^4\Big]
\end{eqnarray}
Hence, $L$ will be
\begin{eqnarray}\label{L}
L&=&{\frac {1}{240}}{\frac {\theta\left( 240{l}^{4}{\rho}^{4}+320
		{l}^{4}{\rho}^{4}{\theta}^{2}+450{\rho}^{8}{\theta}^{4}+360{\theta
		}^{2}{\rho}^{6}{l}^{2}+272{l}^{4}{\rho}^{4}{\theta}^{4}+720{\theta
	}^{4}{l}^{2}{\rho}^{6} \right) }{{l}^{4}{\rho}^{2}}}\nonumber\\
&&+{\frac {1}{240}}
{\frac {\theta\left( -360{\theta}^{2}{\rho}^{3}{l}^{2}{r_+^3}-
		675{\rho}^{5}{\theta}^{4}{r_+^3}-495{\theta}^{4}{\rho}^{3}{l}^{2}{
			r_+^3} \right) \epsilon}{{l}^{4}{\rho}^{2}}}\nonumber\\
&&+ \left( {\frac {1}{240}}
{\frac {\theta\left( 420{r_+^4}{\theta}^{4}{\rho}^{2}{l}^{2}+
		360{r_+^4}{\theta}^{2}{\rho}^{2}{l}^{2}+600{\theta}^{4}{\rho}^{4}{
			r_+^4} \right) }{{l}^{4}{\rho}^{2}}}-{\frac {25}{16}}{\frac {{\theta
		}^{5}{r_+^7}\epsilon}{{l}^{4}\rho}} \right) \delta.
\end{eqnarray}
By integrating we find
\begin{eqnarray}\label{Lint}
\int_0^{\theta_0}Ld\theta&=&-{\frac {1}{1440}}{\frac {{\theta_0^2} \left( 675{\theta_0^4}{\rho}^{4}{r_+^3}+540{\rho}^{2}{\theta_0^2}{r_+^
			3}+375{\theta_0^4}{r_+^7}\delta+495{\theta_0^4}{\rho}^
		{2}{r_+^3} \right) \epsilon}{\rho}}\nonumber\\
&&+\frac {1}{1440}{
	\frac{{\theta_0^2} \left(450{\theta_0^4}{\rho}^{7}+
		272{\theta_0^4}{\rho}^{3}{l}^{4}+720{\theta_0^4}{\rho}
		^{5}+540{\rho}^{5}{\theta_0^2}+480{\rho}^{3}{\theta_0^2}+420{\theta_0^4}\rho\delta{r_+
			^4} \right) }{\rho}}\nonumber\\
&&+\frac{1}{1440}\frac{\theta_0^2\left(+600{\theta_0^4}{\rho}^{3}\delta{r_+^4}+720{\rho}^{3}
	+540\rho{\theta_0^2}\delta{r_+^4}\right)}{\rho}.
\end{eqnarray}
\newpage
Solving the definite integral (\ref{Vcc}), and leaving only the terms in $\theta$ up to order 6th yields
\raggedbottom
\begin{eqnarray}\label{V1}
V_{1}&=&\frac{r_+^3}{4a^3 b^3}\times[2a\epsilon{\rho}^{2}+2a{r_+^2}\delta+2a{r_+^4}\delta+2a{r_+^
	2}\epsilon-2b\epsilon{\rho}^{2}-2b{r_+^2}\epsilon+2a\epsilon+4
a{r_+^4}\delta\epsilon{\rho}^{2}+3a\epsilon{r_+^6}\delta{
	\rho}^{2}\nonumber\\
&&+2a{r_+^2}\epsilon{\rho}^{2}+2a{h}^{2}\delta{\rho}^{2
}+2a{\rho}^{2}\delta{r_+^4}+4a{r_+^4}\delta\epsilon+3a
\epsilon{r_+^6}\delta-3b{r_+^4}\delta\epsilon{\rho}^{2}-3b{r_+
	^6}\delta\epsilon{\rho}^{2}-2br_+\delta\rho\nonumber\\
&&-2\,b{r_+^3}\delta
\rho-2br_+\delta{\rho}^{3}-2b{r_+^3}\delta{\rho}^{3}-4b{r_+^
	4}\delta\epsilon-4b{r_+^6}\delta\epsilon-2b{r_+^2}\epsilon{
	\rho}^{2}+3b{r_+^4}\delta\epsilon a{\rho}^{2}\ln  \left( {\frac {
		1+a}{\rho}} \right)\nonumber\\
&&-3b{r_+^4}\delta\epsilon a\ln  \left( {\frac 
	{1+b}{r_+}} \right) {\rho}^{2}+3b{r_+^6}\delta\epsilon a{\rho}^{2}
\ln  \left( {\frac {1+a}{\rho}} \right) -3b{r_+^6}\delta\epsilon
a\ln  \left( {\frac {1+b}{r_+}} \right) {\rho}^{2}-2b\epsilon\nonumber\\
&&+3b{h}^
{6}\delta\epsilon a\ln  \left( {\frac {1+a}{\rho}} \right)+3b{r_+
	^4}\delta\epsilon a\ln  \left( {\frac {1+a}{\rho}} \right)-3b{r_+
	^4}\delta\epsilon a\ln  \left( {\frac {1+b}{r_+}} \right)-3b{r_+^
	6}\delta\epsilon a\ln  \left( {\frac {1+b}{h}} \right)]
\,\nonumber\\
&&-\frac{1}{8a^5\rho^2}\times[ -\epsilon{r_+^4}\delta{\rho}^{2}+2{\rho}^{3}{\epsilon}^{
	2}{r_+^3}+2r_+\delta{\rho}^{3}+4r_+\delta{\rho}^{5}+2{r_+^5}{
	\delta}^{2}\rho-2\epsilon{\rho}^{4}-4\epsilon{\rho}^{6}-{r_+^4
}\delta{\rho}^{4}\epsilon\nonumber\\
&&-3{\epsilon}^{2}{r_+^7}\delta\rho-2\epsilon{\rho}^{8}+2{r_+^5}{\delta}^{2}{\rho}^{3}+2r_+\delta{
	\rho}^{7}+3\epsilon{r_+^8}{\delta}^{2}+2{\rho}^{5}{\epsilon}^{2}
{r_+^3} ]{r_+^3}{\theta}^{2}
\,\nonumber\\
&&+{\frac {1}{384}}{\frac {{r_+^3}{\theta}^{4}}{{\rho}^{4} \left( 1+a
		\right) ^{2} a^{8}}}\times
[( -320h\delta{\rho}^{7}+424\epsilon{r_+^4}\delta{\rho}
^{10}+108\epsilon{\rho}^{2}{\delta}^{2}{r_+^8}-606{\epsilon}^{2}
{r_+^7}\delta{\rho}^{5}\nonumber\\
&&-798{\epsilon}^{2}{r_+^7}\delta{\rho}^{7}+516\epsilon{r_+^8}{\delta}^{2}{\rho}^{4}+244{r_+^4}\delta{
	\rho}^{4}\epsilon+708\epsilon{r_+^8}{\delta}^{2}{\rho}^{6}+396{
	\epsilon}^{3}{r_+^{10}}\delta{\rho}^{4}+108{\rho}^{6}{\epsilon}^{3}{
	r_+^6}\nonumber\\
&&-36{\rho}^{8}{\epsilon}^{3}{r_+^6}-80r_+\delta{\rho}^{13}-
320r_+\delta{\rho}^{11}-212{\rho}^{5}{\epsilon}^{2}{r_+^3}+288{r_+
	^9}{\delta}^{3}{\rho}^{5}-80{r_+^5}{\delta}^{2}{\rho}^{3}-480r_+
\delta{\rho}^{9}\nonumber\\
&&-96{r_+^5}{\delta}^{2}{\rho}^{5}-176{\rho}^{11}{
	\epsilon}^{2}{r_+^3}-624{\rho}^{7}{\epsilon}^{2}{r_+^3}+84{\rho}^{
	4}{\epsilon}^{3}{r_+^6}-600{\rho}^{9}{\epsilon}^{2}{r_+^3}+378{
	\epsilon}^{3}{r_+^{10}}\delta{\rho}^{6}\nonumber\\
&&-810{\epsilon}^{2}{r_+^{11}}{
	\delta}^{2}{\rho}^{5}+300\epsilon{r_+^8}{\delta}^{2}{\rho}^{8}+560
\epsilon{\rho}^{8}+432{r_+^{12}}{\delta}^{3}\epsilon{\rho}^{4}+
912\epsilon{r_+^4}\delta{\rho}^{8}+792\epsilon{r_+^4}\delta
{\rho}^{6}\,\nonumber\\
&&-900{\epsilon}^{2}{r_+^{11}}{\delta}^{2}{\rho}^{3}+504{r_+
	^{12}}{\delta}^{3}\epsilon{\rho}^{2}-138{\rho}^{3}{\epsilon}^{2}{r_+
	^7}\delta-90{\epsilon}^{2}{r_+^{11}}{\delta}^{2}\rho+18{\epsilon}^
{3}{r_+^{10}}\delta{\rho}^{2}+352{r_+^5}{\delta}^{2}{\rho}^{9}\nonumber\\
&&+192
{r_+^5}{\delta}^{2}{\rho}^{7}-80r_+\delta{\rho}^{5}+72{r_+^9}{
	\delta}^{3}{\rho}^{7}a-30{\rho}^{10}{\epsilon}^{3}{r_+^6}a+6{\rho}
^{13}{\epsilon}^{2}{r_+^3}a+72{r_+^5}{\delta}^{2}{\rho}^{11}a\nonumber\\
&&+216{
	r_+^9}{\delta}^{3}{\rho}^{5}a-80{r_+^5}{\delta}^{2}{\rho}^{3}a-56{
	r_+^5}{\delta}^{2}{\rho}^{5}a-360r_+\delta{\rho}^{9}a+72{r_+^{12}}{
	\delta}^{3}\epsilon a-40r_+\delta{\rho}^{13}a\nonumber\\
&&-200r_+\delta{\rho}^
{11}a-212{\rho}^{5}{\epsilon}^{2}{r_+^3}a-518{\rho}^{7}{\epsilon}^
{2}{r_+^3}a-394{\rho}^{9}{\epsilon}^{2}{r_+^3}a+84{\rho}^{4}{
	\epsilon}^{3}{r_+^6}a+66{\rho}^{6}{\epsilon}^{3}{r_+^6}a\nonumber\\
&&-48{\rho}^
{8}{\epsilon}^{3}{r_+^6}a-82{\rho}^{11}{\epsilon}^{2}{r_+^3}a-80r_+
\delta{\rho}^{5}a-280r_+\delta{\rho}^{7}a+200{r_+^5}{\delta}^{2}
{\rho}^{7}a+248{r_+^5}{\delta}^{2}{\rho}^{9}a\nonumber\\
&&+144{r_+^9}{\delta}^{
	3}{\rho}^{3}a+144{r_+^9}{\delta}^{3}{\rho}^{3}+60\epsilon{r_+^4}
\delta{\rho}^{12}+128\epsilon{\rho}^{6}+24\epsilon{\rho}^{16
}a+536\epsilon{\rho}^{12}a+184\epsilon{\rho}^{14}a\nonumber\\
&&+128
\epsilon{\rho}^{6}a+496\epsilon{\rho}^{8}a+744\epsilon{\rho}
^{10}a+72{r_+^12}{\delta}^{3}\epsilon+144{r_+^9}{\delta}^{3}{\rho}
^{7}-60\,{\rho}^{10}{\epsilon}^{3}{r_+^6}+12{\rho}^{13}{\epsilon}^{2
}{r_+^3}\nonumber\\
&&+144{r_+^5}{\delta}^{2}{\rho}^{11}+242\epsilon{r_+^4}
\delta{\rho}^{10}a+108\epsilon{\rho}^{2}{\delta}^{2}{r_+^8}a-537
{\epsilon}^{2}{r_+^7}\delta{\rho}^{5}a-564{\epsilon}^{2}{r_+^7}
\delta{\rho}^{7}a+462\epsilon{r_+^8}{\delta}^{2}{\rho}^{4}a\nonumber\\
&&+504
\epsilon{r_+^8}{\delta}^{2}{\rho}^{6}a+387{\epsilon}^{3}{r_+^{10}}
\delta{\rho}^{4}a+468{r_+^{12}}{\delta}^{3}\epsilon{\rho}^{2}a-138
{\rho}^{3}{\epsilon}^{2}{r_+^7}\delta a+670\epsilon{r_+^4}
\delta{\rho}^{6}a+244\epsilon{r_+^4}\delta{\rho}^{4}a\nonumber\\
&&-855{
	\epsilon}^{2}{r_+^{11}}{\delta}^{2}{\rho}^{3}a+638\epsilon{r_+^4}
\delta{\rho}^{8}a+18{\epsilon}^{3}{r_+^{10}}\delta\,a{\rho}^{2}-90
{\epsilon}^{2}{r_+^{11}}{\delta}^{2}a\rho-165{\epsilon}^{2}{r_+^7}
\delta{\rho}^{9}a\nonumber\\
&&+150\epsilon{r_+^8}{\delta}^{2}{\rho}^{8}a+189
{\epsilon}^{3}{r_+^{10}}\delta{\rho}^{6}a+216{r_+^{12}}{\delta}^{3}
\epsilon{\rho}^{4}a-405{\epsilon}^{2}{r_+^{11}}{\delta}^{2}{\rho}^{5
}a+30\epsilon{r_+^4}\delta{\rho}^{12}a+320\,\epsilon\,{\rho}^{14
}\nonumber\\
&&+960\epsilon{\rho}^{10}+48\epsilon{\rho}^{16}+800\epsilon
{\rho}^{12}-330{\epsilon}^{2}{r_+^7}\delta{\rho}^{9}) {r_+^
	3}{\theta}^{4}]\,.
\end{eqnarray}
Now, by taking series in $\theta$ (up to order 4),  and leaving only linear terms in $\epsilon$ and $\delta$, we find
\begin{eqnarray}\label{V4}
V_4&=&\frac{1}{24a^3 b^3}\nonumber\\
&&\times( -3\,{\theta}^{2}{r_+^6}\delta\,\epsilon\,ba\ln  \left( 1+a
\right) {\rho}^{2}-3\,{\theta}^{2}{r_+^4}\delta\,\epsilon\,ba\ln 
\left( 1+a \right) {\rho}^{2}+3\,{\theta}^{2}{r_+^6}\delta\,\epsilon
\,ba\ln  \left( 1+b \right) {\rho}^{2}\nonumber\\
&&+3\,{\theta}^{2}{r_+^6}\delta\,
\epsilon\,ba\ln  \left( \rho \right) {\rho}^{2}-3\,{\theta}^{2}{r_+^4}
\delta\,\epsilon\,ba\ln  \left( r_+ \right) {\rho}^{2}+3\,{\theta}^{2}{r_+
	^4}\delta\,\epsilon\,ba\ln  \left( 1+b \right) {\rho}^{2}\nonumber\\
&&-3\,{\theta
}^{2}{r_+^6}\delta\,\epsilon\,ba\ln  \left( r_+ \right) {\rho}^{2}+3\,{
\theta}^{2}{r_+^4}\delta\,\epsilon\,ba\ln  \left( \rho \right) {\rho}^
{2}+12\,\epsilon\,a-12\,\epsilon\,b-18\,{r_+^4}\delta\,\epsilon\,ba{
	\rho}^{2}\ln  \left( 1+b \right)\nonumber\\
&& +18\,{r_+^4}\delta\,\epsilon\,ba{\rho
}^{2}\ln  \left( r_+ \right)-18\,{r_+^6}\delta\,\epsilon\,ba{\rho}^{2}
\ln  \left( 1+b \right) +18\,{r_+^6}\delta\,\epsilon\,ba{\rho}^{2}\ln 
\left( r_+ \right)\nonumber\\
&& +18\,{r_+^6}\delta\,\epsilon\,ba{\rho}^{2}\ln 
\left( 1+a \right) -3\,{\theta}^{2}{r_+^6}\delta\,\epsilon\,ba\ln 
\left( 1+a \right) -3\,{\theta}^{2}{r_+^4}\delta\,\epsilon\,ba\ln 
\left( 1+a \right) \nonumber\\
&&+3\,{\theta}^{2}{r_+^6}\delta\,\epsilon\,ba\ln 
\left( 1+b \right) +3\,{\theta}^{2}{r_+^6}\delta\,\epsilon\,ba\ln 
\left( \rho \right) -3\,{\theta}^{2}{r_+^4}\delta\,\epsilon\,ba\ln 
\left( r_+ \right) \nonumber\\
&&+3\,{\theta}^{2}{r_+^4}\delta\,\epsilon\,ba\ln 
\left( 1+b \right) -3\,{\theta}^{2}{r_+^6}\delta\,\epsilon\,ba\ln 
\left( r_+ \right) +3\,{\theta}^{2}{r_+^4}\delta\,\epsilon\,ba\ln 
\left( \rho \right)-18\,{r_+^6}\delta\,\epsilon\,ba{\rho}^{2}\ln 
\left( \rho \right) \nonumber\\
&&+18\,{r_+^4}\delta\,\epsilon\,ba{\rho}^{2}\ln 
\left( 1+a \right) -18\,{r_+^4}\delta\,\epsilon\,ba{\rho}^{2}\ln 
\left( \rho \right) +18\,{r_+^4}\delta\,\epsilon\,ba\ln  \left( r_+
\right)-18\,{r_+^6}\delta\,\epsilon\,ba\ln  \left( 1+b \right)\nonumber\\
&&+18\,
{r_+^6}\delta\,\epsilon\,ba\ln  \left( r_+ \right) +18\,{r_+^4}\delta\,
\epsilon\,ba\ln  \left( 1+a \right) -18\,{r_+^4}\delta\,\epsilon\,ba
\ln  \left( \rho \right) +18\,{r_+^6}\delta\,\epsilon\,ba\ln  \left( 1
+a \right)\nonumber\\
&& -18\,{r_+^6}\delta\,\epsilon\,ba\ln  \left( \rho \right) +2
\,{\theta}^{2}{r_+^3}\delta\,b{\rho}^{3}+4\,{\theta}^{2}{r_+^6}\delta
\,\epsilon\,b+3\,{\theta}^{2}{r_+^6}\delta\,\epsilon\,b{\rho}^{2}+4\,{
	\theta}^{2}{r_+^4}\delta\,\epsilon\,b+3\,{\theta}^{2}{r_+^4}\delta\,
\epsilon\,b{\rho}^{2}\nonumber\\
&&-4\,{\theta}^{2}{r_+^4}\delta\,\epsilon\,a-4\,{
	\theta}^{2}{r_+^4}\delta\,\epsilon\,a{\rho}^{2}-3\,{\theta}^{2}{r_+^6}
\delta\,\epsilon\,a-3\,{\theta}^{2}{r_+^6}\delta\,\epsilon\,a{\rho}^{2
}-2\,{\theta}^{2}\epsilon\,a{r_+^2}{\rho}^{2}+2\,{\theta}^{2}\epsilon
\,b{r_+^2}{\rho}^{2}\nonumber\\
&&-2\,{\theta}^{2}{r_+^2}\delta\,a{\rho}^{2}-2\,{
	\theta}^{2}{r_+^4}\delta\,a{\rho}^{2}+2\,{\theta}^{2}r_+ \delta\,b\rho+2
\,{\theta}^{2}{r_+^3}\delta\,b\rho-18\,{r_+^4}\delta\,\epsilon\,ba\ln 
\left( 1+b \right) +6\,{\theta}^{2}{\rho}^{4}\epsilon\,b\nonumber\\
&&+6\,{\theta}^{2}{\rho}^{2}\epsilon\,b-6\,{\theta}^{
	2}{r_+^3}\delta\,b\rho-6\,{\theta}^{2}r_+ \delta\,b{\rho}^{3}-6\,{\theta}^{2}{r_+^3}\delta\,b{
	\rho}^{3}+6\,{\theta}^{2}{\rho}^{2}\epsilon\,b{r_+^2}+
6\,{\theta}^{2}{\rho}^{4}\epsilon\,b{r_+^2}+3\,{\theta
}^{2}{r_+^4}\delta\,b\epsilon\nonumber\\
&&+3\,{\theta}^{2}{r_+^6}
\delta\,b\epsilon-6\,{\theta}^{2}r_+ \delta\,b\rho+2\,{\theta}^{2}r_+ \delta\,b{\rho}^{3}+2\,{\theta}^{2}\epsilon
\,b{\rho}^{2}+2\,{\theta}^{2}\epsilon\,b{r_+^2}-2\,{\theta}^{2}
\epsilon\,a{r_+^2}-2\,{\theta}^{2}\epsilon\,a{\rho}^{2}\nonumber\\
&&-2\,{\theta}^{2
}{r_+^4}\delta\,a-2\,{\theta}^{2}{r_+^2}\delta\,a-18\,{r_+^4}\delta\,
\epsilon\,b{\rho}^{2}-18\,{r_+^6}\delta\,\epsilon\,b{\rho}^{2}+24\,{r_+^4}\delta\,\epsilon\,a{\rho}^{2}+18\,{r_+^6}\delta\,\epsilon\,a{\rho}
^{2}-2\,{\theta}^{2}\epsilon\,a\nonumber\\
&&+2\,{\theta}^{2}\epsilon\,b+12\,
\epsilon\,a{r_+^2}{\rho}^{2}-12\,\epsilon\,b{r_+^2}{\rho}^{2}+12\,{r_+^
	2}\delta\,a{\rho}^{2}+12\,{r_+^4}\delta\,a{\rho}^{2}-12\,r_+ \delta\,b
\rho-12\,{r_+^3}\delta\,b\rho\nonumber\\
&&-12\,r_+ \delta\,b{\rho}^{3}-12\,{r_+^3}
\delta\,b{\rho}^{3}-24\,{r_+^6}\delta\,\epsilon\,b-24\,{r_+^4}\delta\,
\epsilon\,b+24\,{r_+^4}\delta\,\epsilon\,a+18\,{r_+^6}\delta\,\epsilon
\,a-12\,\epsilon\,b{\rho}^{2}\nonumber\\
&&-12\,\epsilon\,b{r_+^2}+12\,\epsilon\,a{r_+
	^2}+12\,\epsilon\,a{\rho}^{2}+12\,{r_+^4}\delta\,a+12\,{r_+^2}\delta
\,a ) {r_+^3}\theta\,,
\end{eqnarray}
where we have defined $V_4=V_3\sin(\theta)$. Under these expansions, the complexity volume becomes\newpage
\begin{eqnarray}\label{Vccc}
V_c=2\pi \int_0^{\theta_0}V4 d\theta&=&\frac{1}{48a^3b^3}\times( 6\,{\theta_0^2}{\rho}^{4}\epsilon\,b+6\,{\theta_
	0^2}{\rho}^{2}\epsilon\,b+3\,{\theta_0^2}{
	r_+^4}\delta\,\epsilon\,ba{\rho}^{2}\ln  \left( 1+b \right) \nonumber\\
&&+3\,{
	\theta_0^2}{r_+^4}\delta\,\epsilon\,ba{\rho}^{2}\ln  \left( \rho
\right)+3\,{\theta_0^2}{r_+^6}\delta\,\epsilon\,ba{\rho}^{2}
\ln  \left( \rho \right) -3\,{\theta_0^2}{r_+^4}\delta\,\epsilon
\,ba{\rho}^{2}\ln  \left( 1+a \right)\nonumber\\
&& +3\,{\theta_0^2}{r_+^6}
\delta\,\epsilon\,ba{\rho}^{2}\ln  \left( 1+b \right) -3\,{\theta_0^2}{r_+^6}\delta\,\epsilon\,ba{\rho}^{2}\ln  \left( r_+ \right) -3\,{
	\theta_0^2}{r_+^4}\delta\,\epsilon\,ba{\rho}^{2}\ln  \left( r_+
\right) \nonumber\\
&&-3\,{\theta_0^2}{r_+^6}\delta\,\epsilon\,ba{\rho}^{2}
\ln  \left( 1+a \right) +6\,{\theta_0^2}{\rho}^{2}\epsilon\,
b{r_+^2}+6\,{\theta_0^2}{\rho}^{4}\epsilon\,
b{r_+^2}+3\,{\theta_0^2}{r_+^4}\delta\,b\epsilon\nonumber\\
&&+3\,{\theta_0^2}{r_+^6}\delta\,b
\epsilon-6\,{\theta_0^2}r_+\delta\,b\rho-6\,{
	\theta_0^2}{r_+^3}\delta\,b\rho-6\,{\theta_0^2}r_+\delta\,b{\rho}^{3}-6\,{\theta_0^2}{r_+^3}
\delta\,b{\rho}^{3}\nonumber\\
&&-3\,{\theta_0^2}{r_+^6}\delta
\,\epsilon\,ba\ln  \left( 1+a \right) +3\,{\theta_0^2}{r_+^4}
\delta\,\epsilon\,ba\ln  \left( 1+b \right) -3\,{\theta_0^2}{r_+^
	4}\delta\,\epsilon\,ba\ln  \left( r_+ \right)\nonumber\\
&& +36\,{r_+^4}\delta\,
\epsilon\,ba{\rho}^{2}\ln  \left( r_+ \right) -36\,{r_+^6}\delta\,
\epsilon\,ba{\rho}^{2}\ln  \left( 1+b \right) +36\,{r_+^6}\delta\,
\epsilon\,ba{\rho}^{2}\ln  \left( r_+ \right) \nonumber\\
&&+36\,{r_+^6}\delta\,
\epsilon\,ba{\rho}^{2}\ln  \left( 1+a \right) -36\,{r_+^6}\delta\,
\epsilon\,ba{\rho}^{2}\ln  \left( \rho \right) +36\,{r_+^4}\delta\,
\epsilon\,ba{\rho}^{2}\ln  \left( 1+a \right) \nonumber\\
&&-36\,{r_+^4}\delta\,
\epsilon\,ba{\rho}^{2}\ln  \left( \rho \right) -36\,{r_+^4}\delta\,
\epsilon\,ba{\rho}^{2}\ln  \left( 1+b \right) +24\,\epsilon\,a{r_+^2}{
	\rho}^{2}-24\,\epsilon\,b{r_+^2}{\rho}^{2}\nonumber\\
&&+24\,{r_+^2}\delta\,a{\rho}^
{2}+24\,{r_+^4}\delta\,a{\rho}^{2}-24\,r_+\delta\,b\rho-24\,{r_+^3}
\delta\,b\rho-24\,r_+\delta\,b{\rho}^{3}-24\,{r_+^3}\delta\,b{\rho}^{3}\nonumber\\
&&-48\,{r_+^6}\delta\,\epsilon\,b-48\,{r_+^4}\delta\,\epsilon\,b+48\,{r_+^
	4}\delta\,\epsilon\,a+36\,{r_+^6}\delta\,\epsilon\,a-2\,{\theta_0^2}\epsilon\,a-2\,{\theta_0^2}\epsilon\,a{\rho}^{2}+2\,{\theta
	_0^2}\epsilon\,b\nonumber\\
&&+2\,{\theta_0^2}\epsilon\,b{\rho}^{2}-2\,{
	\theta_0^2}\epsilon\,a{r_+^2}{\rho}^{2}+2\,{\theta_0^2}
\epsilon\,b{r_+^2}{\rho}^{2}-2\,{\theta_0^2}{r_+^2}\delta\,a{
	\rho}^{2}-2\,{\theta_0^2}{r_+^4}\delta\,a{\rho}^{2}\nonumber\\
&&+2\,{\theta_0^2}r_+\delta\,b\rho+2\,{\theta_0^2}{r_+^3}\delta\,b\rho+4\,{
	\theta_0^2}{r_+^6}\delta\,\epsilon\,b+4\,{\theta_0^2}{r_+^
	4}\delta\,\epsilon\,b-4\,{\theta_0^2}{r_+^4}\delta\,\epsilon\,a-
3\,{\theta_0^2}{h}^{6}\delta\,\epsilon\,a\nonumber\\
&&-36\,{r_+^4}\delta\,
\epsilon\,ba\ln  \left( 1+b \right) +36\,{r_+^4}\delta\,\epsilon\,ba
\ln  \left( r_+ \right) -36\,{r_+^6}\delta\,\epsilon\,ba\ln  \left( 1+b
\right)\nonumber\\
&& +36\,{r_+^6}\delta\,\epsilon\,ba\ln  \left( r_+ \right) +36\,{r_+
	^4}\delta\,\epsilon\,ba\ln  \left( 1+a \right) -36\,{r_+^4}\delta\,
\epsilon\,ba\ln  \left( \rho \right) \nonumber\\
&&+36\,{r_+^6}\delta\,\epsilon\,ba
\ln  \left( 1+a \right) -36\,{r_+^6}\delta\,\epsilon\,ba\ln  \left( 
\rho \right) -24\,\epsilon\,b{\rho}^{2}-24\,\epsilon\,b{r_+^2}+24\,
\epsilon\,a{r_+^2}\nonumber\\
&&+24\,\epsilon\,a{\rho}^{2}+24\,{r_+^4}\delta\,a+24\,
{r_+^2}\delta\,a+24\,\epsilon\,a-24\,\epsilon\,b-36\,{r_+^4}\delta\,
\epsilon\,b{\rho}^{2}-36\,{r_+^6}\delta\,\epsilon\,b{\rho}^{2}\nonumber\\
&&+48\,{r_+
	^4}\delta\,\epsilon\,a{\rho}^{2}+36\,{r_+^6}\delta\,\epsilon\,a{\rho}
^{2}+2\,{\theta_0^2}r_+\delta\,b{\rho}^{3}+2\,{\theta_0^2}{r_+
	^3}\delta\,b{\rho}^{3}+2\,{\theta_0^2}\epsilon\,b{r_+^2}-2\,{
	\theta_0^2}\epsilon\,a{r_+^2}\nonumber\\
&&-2\,{\theta_0^2}{r_+^4}\delta
\,a-2\,{\theta_0^2}{r_+^2}\delta\,a+3\,{\theta_0^2}{r_+^4}
\delta\,\epsilon\,b{\rho}^{2}+3\,{\theta_0^2}{r_+^6}\delta\,
\epsilon\,b{\rho}^{2}-4\,{\theta_0^2}{r_+^4}\delta\,\epsilon\,a{
	\rho}^{2}\nonumber\\
&&-3\,{\theta_0^2}{r_+^6}\delta\,\epsilon\,a{\rho}^{2}-3
\,{\theta_0^2}{r_+^6}\delta\,\epsilon\,ba\ln  \left( r_+ \right) +
3\,{\theta_0^2}{r_+^6}\delta\,\epsilon\,ba\ln  \left( \rho
\right) \nonumber\\
&&+3\,{\theta_0^2}{r_+^4}\delta\,\epsilon\,ba\ln  \left( 
\rho \right) -3\,{\theta_0^2}{r_+^4}\delta\,\epsilon\,ba\ln 
\left( 1+a \right) +3\,{\theta_0^2}{r_+^6}\delta\,\epsilon\,ba
\ln  \left( 1+b \right) 
{\theta_0^2}{r_+^3}\pi
) \nonumber\\
\end{eqnarray}
\section{SAdS for any dimension}\label{sAdsanyd}
\begin{eqnarray}
P_{fid}&=&-\frac{\partial M}{\partial V_{Fid}}=-\frac{\frac{\partial M}{\partial r_{+}}}{\frac{\partial V_{Fid}}{\partial r_{+}}}=
24\, \left( -{r_+^2}-3\,{r_+^4}+{Q}^{2} \right)  b^5 \left( 1+b \right)\nonumber\\
&&\times({r_+^4}{\theta_0^2}\pi \, ( 252\,{r_+^8}\delta\,\epsilon+
372\,\delta\,{r_+^4}\epsilon-8\,{r_+^6}{\theta_0^2}\delta-48\,{r_+
	^5}{\theta_0^2}\delta+18\,{\theta_0^2}\epsilon\,\rho-16
\,{\theta_0^2}r_+\delta-48\,{\theta_0^2}{r_+^3}\delta\nonumber\\
&&-16\,{r_+
	^7}{\theta_0^2}\delta+18\,{\theta_0^2}\epsilon\,\rho\,{r_+
	^6}+216\,\delta\,{r_+^4}-6\,{\theta_0^2}\epsilon+120\,\epsilon
\,{r_+^2}+120\,{r_+^2}\delta+96\,{r_+^6}\delta-96\,r_+\delta+48\,{r_+^4}
\epsilon\nonumber\\
&&-288\,{r_+^5}\delta-96\,\delta\,{r_+^7}-288\,\delta\,{r_+^3}-
21\,{r_+^8}b{\theta_0^2}\delta\,\epsilon+21\,{r_+^{10}}{\theta_0^2}\delta\,\epsilon\,\ln  \left( {\frac {1+b}{r_+}} \right) \nonumber\\
&&+63\,{r_+^6}{\theta_0^2}\delta\,\epsilon\,\ln  \left( {\frac {1+b}{r_+}}
\right)+63\,{r_+^8}{\theta_0^2}\delta\,\epsilon\,\ln  \left( {
	\frac {1+b}{r_+}} \right) +21\,{r_+^4}{\theta_0^2}\delta\,\epsilon
\,\ln  \left( {\frac {1+b}{r_+}} \right) \nonumber\\
&&-504\,{r_+^6}\delta\,\epsilon\,
b\ln  \left( {\frac {1+b}{r_+}} \right) -252\,\delta\,{r_+^4}\epsilon\,b
\ln  \left( {\frac {1+b}{r_+}} \right)+18\,{r_+^4}{\theta_0^2}
\epsilon\,\rho\,b-252\,{r_+^8}\delta\,\epsilon\,b\ln  \left( {\frac {1
		+b}{r_+}} \right) \nonumber\\
&&+36\,{\theta_0^2}\epsilon\,{r_+^2}\rho\,b-31\,{r_+
	^4}{\theta_0^2}\delta\,\epsilon\,b-49\,{r_+^6}{\theta_0^
	2}\delta\,\epsilon\,b-6\,{\theta_0^2}\epsilon\,b+120\,\epsilon\,
b{r_+^2}-96\,r_+\delta\,b-192\,\delta\,{r_+^3}b\nonumber\\
&&-96\,{r_+^5}b\delta+48\,{
	r_+^4}\epsilon\,b+96\,{r_+^6}\delta\,b+216\,\delta\,{r_+^4}b+120\,{r_+^
	2}\delta\,b-756\,{r_+^6}\delta\,\epsilon\,\ln  \left( {\frac {1+b}{r_+}
} \right) \nonumber\\
&&-252\,\delta\,{r_+^4}\epsilon\,\ln  \left( {\frac {1+b}{r_+}}
\right) -756\,{r_+^8}\delta\,\epsilon\,\ln  \left( {\frac {1+b}{r_+}}
\right)-252\,{r_+^{10}}\delta\,\epsilon\,\ln  \left( {\frac {1+b}{r_+}}
\right) -16\,{\theta_0^2}h\delta\,b\nonumber\\
&&-32\,{\theta_0^2}{r_+^
	3}\delta\,b-10\,{\theta_0^2}\epsilon\,b{r_+^2}+252\,{r_+^8}
\delta\,\epsilon\,b-8\,{r_+^6}{\theta_0^2}\delta\,b-4\,{r_+^4}{
	\theta_0^2}\epsilon\,b-18\,{r_+^4}{\theta_0^2}\delta\,b-16
\,{r_+^5}b{\theta_0^2}\delta\nonumber\\
&&-10\,{\theta_0^2}{r_+^2}\delta
\,b+18\,{\theta_0^2}\epsilon\,b\rho+588\,{r_+^6}\delta\,\epsilon
\,b+372\,\delta\,{r_+^4}\epsilon\,b+72\,\epsilon\,b+72\,\epsilon+54\,{
	r_+^4}{\theta_0^2}\epsilon\,\rho+54\,{\theta_0^2}\epsilon
\,{r_+^2}\rho\nonumber\\
&&-49\,{r_+^6}{\theta_0^2}\delta\,\epsilon-21\,{r_+^8
}{\theta_0^2}\delta\,\epsilon-31\,{r_+^4}{\theta_0^2}
\delta\,\epsilon-10\,{\theta_0^2}\epsilon\,{r_+^2}-18\,{r_+^4}{
	\theta_0^2}\delta-4\,{r_+^4}{\theta_0^2}\epsilon-10\,{
	\theta_0^2}{r_+^2}\delta\nonumber\\
&&+588\,{r_+^6}\delta\,\epsilon+21\,{r_+^4
}{\theta_0^2}\delta\,\epsilon\,b\ln  \left( {\frac {1+b}{r_+}}
\right) +42\,{r_+^6}{\theta_0^2}\delta\,\epsilon\,b\ln  \left( 
{\frac {1+b}{r_+}} \right) +21\,{r_+^8}b{\theta_0^2}\delta\,
\epsilon\,\ln  \left( {\frac {1+b}{r_+}} \right)  ) )
^{-1}.\nonumber\\
\end{eqnarray}
Because we are interested to small deformations of the AdS background, 
taking series in $\epsilon$ and $\theta$, and leaving only linear terms in them:

\begin{eqnarray}\label{P3}
P&=&12\left( -{r_+^2}-3\,{r_+^4}+{Q}^{2} \right)  b^{5} \left( 1+b \right)  ( 72+120\,{r_+
	^2}+48\,{r_+^4}+120\,b{r_+^2}+48\,b{r_+
	^4}-10\,{\theta_0^2}{r_+^2}\nonumber\\
&&+18\,{\theta_0^2}{r_+^4}\rho
\,b+54\,{\theta_0^2}\rho\,{r_+^2}+54\,{\theta_0^2}{r_+^4}\rho+36\,{\theta_0^2}\rho\,{r_+^2}b
-6\,{\theta_0^2}-10\,{\theta_0^2}b{r_+^2}
-4\,{\theta_0^2}{r_+^4}b+18\,{\theta_0^2}
\rho\,{r_+^6}\nonumber\\
&&-6\,{\theta_0^2}b+72\,b+18\,{\theta_0^2}b\rho+18\,{\theta_0^2
}\rho-4\,{\theta_0^2}{r_+^4} ) \epsilon\nonumber\\
&&\times({r_+^5}{\theta_0^2}\pi \, ( 48+48\,b+9\,{\theta_0^2}{
	h}^{3}+5\,{\theta_0^2}h+24\,{r_+^4}{\theta_0^2}+4\,{r_+^5}
{\theta_0^2}+8\,{\theta_0^2}{r_+^6}+24\,{\theta_0^2}
{r_+^2}+8\,{\theta_0^2}\nonumber\\
&&-48\,{r_+^5}+48\,{r_+^6}+144\,{r_+^4}-108
\,{r_+^3}+144\,{r_+^2}-60\,r_+ +96\,b{r_+^2}+48\,b{r_+^4}+8\,{\theta_0^2}b-108\,{r_+^3}b\nonumber\\
&&-48\,{r_+^5}b-60\,r_+ b+16\,{\theta_0^2}b{r_+^
	2}+8\,{r_+^4}{\theta_0^2}b+5\,{\theta_0^2}r_+ b+9\,{\theta_0^2}{r_+^3}b+4\,{r_+^5}b{\theta_0^2} )  ( -48\,{
	\theta_0^2}{r_+^3}-16\,{\theta_0^2}r_+ \nonumber\\
&&-18\,{r_+^4}{\theta_0^2}-48\,{r_+^5}{\theta_0^2}-8\,{\theta_0^2}{r_+^6}-10
\,{\theta_0^2}{r_+^2}-16\,{r_+^7}{\theta_0^2}-288\,{r_+^5}
-96\,{r_+^7}+96\,{r_+^6}+216\,{r_+^4}\nonumber\\
&&-288\,{r_+^3}+120\,{r_+^2}-96\,r_+ +
120\,b{r_+^2}+216\,b{r_+^4}-192\,{r_+^3}b-96\,{r_+^5}b-96\,r_+ b+96\,b{r_+^6}\nonumber\\
&&-10\,{\theta_0^2}b{r_+^2}-18\,{r_+^4}{\theta_0^2}b-16
\,{\theta_0^2}r_+ b-32\,{\theta_0^2}{r_+^3}b-16\,{r_+^5}b{
	\theta_0^2}-8\,{\theta_0^2}{r_+^6}b ) {\delta}^{2})
^{-1}\nonumber\\
&&+\frac{24\, \left( -{r_+^2}-3\,{r_+^4}+{Q}^{2} \right)  b^{5} \left( 1+b \right)}{\delta}\nonumber\\
&&\times({r_+^4}{\theta_0^2}\pi \, ( -48\,{\theta_0^2}{r_+^3}-
16\,{\theta_0^2}r_+ -18\,{r_+^4}{\theta_0^2}-48\,{r_+^5}{
	\theta_0^2}-8\,{\theta_0^2}{r_+^6}-10\,{\theta_0^2}{
	r_+^2}-16\,{r_+^7}{\theta_0^2}\nonumber\\
&&-288\,{r_+^5}-96\,{r_+^7}+96\,{r_+^
	6}+216\,{r_+^4}-288\,{r_+^3}+120\,{r_+^2}-96\,r_+ +120\,b{r_+^2}+216\,b{
	r_+^4}-192\,{r_+^3}b\nonumber\\
&&-96\,{r_+^5}b-96\,r_+ b+96\,b{r_+^6}-10\,{\theta_0^2}b{r_+^2}-18\,{r_+^4}{\theta_0^2}b-16\,{\theta_0^2}r_+ b
-32\,{\theta_0^2}{r_+^3}b-16\,{r_+^5}b{\theta_0^2}\nonumber\\
&&-8\,{
	\theta_0^2}{r_+^6}b ) 
)^{-1}\nonumber\\
&&+12\left( -{r_+^2}-3\,{r_+^4}+{Q}^{2} \right)  b^5 \left( 1+b \right) ( -31\,{r_+^4
}{\theta_0^2}-49\,{\theta_0^2}{r_+^6}-21\,{r_+^8}{\theta_0^2}+588\,{r_+^6}+252\,{r_+^8}\nonumber\\
&&+372\,{r_+^4}-252\,{r_+^4}\ln 
\left( {\frac {1+b}{r_+}} \right) +252\,{r_+^8}b-756\,{r_+^6}\ln  \left( {\frac {1+b}{r_+}}
\right)-756\,{r_+^8}\ln  \left( {\frac {1+b}{r_+}}
\right) \nonumber\\
&&-252\,{r_+^10}\ln  \left( {\frac {1+b}{r_+}}
\right) +21\,{r_+^8}b{\theta_0^2}\ln  \left( {
	\frac {1+b}{r_+}} \right)+21\,{\theta_0^2}{r_+^4}
b\ln  \left( {\frac {1+b}{r_+}} \right) 
+42\,{\theta_0^2}{r_+^6}b\ln  \left( {\frac {1+
		b}{r_+}} \right)\nonumber\\
&& +63\,{r_+^6}{\theta_0^2}\ln 
\left( {\frac {1+b}{r_+}} \right)+63\,{r_+^8}{\theta_0^2}\ln  \left( {\frac {1+b}{r_+}} \right) -252\,{r_+
	^4}b\ln  \left( {\frac {1+b}{r_+}}
\right) -252\,{r_+^8}b\ln  \left( {\frac {1+b}{r_+}} \right) \nonumber\\
&&+21\,{r_+^10}{\theta_0^2}\ln  \left( {
	\frac {1+b}{r_+}} \right) -504\,{r_+^6}b
\ln  \left( {\frac {1+b}{r_+}} \right) +21\,{r_+^4}{
	\theta_0^2}\ln  \left( {\frac {1+b}{r_+}} \right)-21\,{r_+^8}b{\theta_0^2}-31\,{r_+^4}{\theta_0^2}b\nonumber\\
&&-49\,{\theta_0^2}{r_+^6}b
+588\,b{r_+^6}+372\,b{r_+^4} ) 
\epsilon\nonumber\\
&&\times({r_+^5}{\theta_0^2}\pi \, ( 48+9\,{\theta_0^2}{r_+^3}
+5\,{\theta_0^2}r_+ +24\,{r_+^4}{\theta_0^2}+4\,{r_+^5}{
	\theta_0^2}+8\,{\theta_0^2}{r_+^6}+24\,{\theta_0^2}{
	r_+^2}+48\,{r_+^6}+144\,{r_+^4}\nonumber\\
&&-108\,{r_+^3}+8\,{\theta_0^2}-48
\,{r_+^5}-60\,r_+ +144\,{r_+^2}+16\,{\theta_0^2}b{r_+
	^2}+8\,{r_+^4}{\theta_0^2}b+9\,{\theta_0^2}{r_+^3}b+5\,{\theta_0^2}r_+ b\nonumber\\
&&+4
\,{r_+^5}b{\theta_0^2}-108\,{r_+^3}b-48\,{r_+^5}b-60\,r_+ b+48\,b+96\,b{r_+^2}+48\,b{r_+^4}+8\,
{\theta_0^2}b )  ( -48\,{\theta_0^2}{r_+^3}-16\,{\theta_0^2}r_+ \nonumber\\
&&-18\,{r_+^4}{\theta_0^2}-48
\,{r_+^5}{\theta_0^2}-8\,{\theta_0^2}{r_+^6}-10\,{\theta_0^2}{r_+^2}-16\,{r_+^7}{\theta_0^2}+96\,{r_+^6}+216\,{r_+^4
}-288\,{r_+^3}-288\,{r_+^5}\nonumber\\
&&-96\,{r_+^7}-96\,r_+ +120\,{r_+^2}-10\,{\theta
	_0^2}b{r_+^2}-18\,{r_+^4}{\theta_0^2}
b-32\,{\theta_0^2}{r_+^3}b-16\,{
	\theta_0^2}r_+ b-16\,{r_+^5}b{
	\theta_0^2}-8\,{\theta_0^2}{r_+^6}b\nonumber\\
&&-192\,{
	r_+^3}b-96\,{r_+^5}b-96\,r_+ b+96\,b{r_+^6}+120\,b{r_+^2}+216\,
b{r_+^4} ) )^{-1}.
\end{eqnarray}

The above expression is the equation  for fidelity susceptibility pressure and volume in our model.

 \end{document}